\documentclass[showkeys,
reprint,
 amsmath,amssymb,
 aps,
]{revtex4-2}

\usepackage[colorlinks=true, linkcolor=blue, urlcolor=blue, citecolor=blue]{hyperref}
\usepackage{graphicx}
\usepackage{dcolumn}
\usepackage{bm}
\usepackage{float}
\usepackage[T1]{fontenc}
\def\aap{Astronomy \& Astrophysics}
\def\apj{Astrophysical Journal}
\def\mnras{Monthly Notices of the RAS}

\def\jgr{Journal of Geophysical Research}
\def\apjl{Astrophysical Journal Letters}
\def\pasj{Publications of the Astronomical Society of Japan}
\def\apjs{Astrophysical Journal Supplement Series}
\def\approxprop{%
  \def\p{%
    \setbox0=\vbox{\hbox{$\propto$}}%
    \ht0=0.6ex \box0 }%
  \def\s{%
    \vbox{\hbox{$\sim$}}%
  }%
  \mathrel{\raisebox{0.7ex}{%
      \mbox{$\underset{\s}{\p}$}%
    }}%
}

\begin{document}

\preprint{APS/123-QED}

\title{Effect of Magnetic Field on the Formation of Radiatively Inefficient Accretion Flow around Black Holes}

\author{Anish Sarkar$^{1,}$}
 \email{sarkar.anish.1001@gmail.com}
\author{Mayukh Pahari$^{1,}$}
 \email{mayukh@phy.iith.ac.in}
\affiliation{Department of Physics, Indian Institute of Technology, Hyderabad, Kandi, Sangareddy, India}%


\date{\today}

\begin{abstract}
We study the effects of magnetic field in the formation of a radiatively inefficient accretion flow (RIAF) in the presence of Bremsstrahlung cooling, which facilitates the formation of a geometrically thin, optically thick accretion disk surrounded by a hot corona. We have performed axis-symmetric magnetohydrodynamic (MHD) simulations of an initial accretion torus with a $1/r$ dependant local poloidal field in the presence of a pseudo-Newtonian potential, taking into account optically thin cooling, resistivity and viscosity. We observe the formation of persistent jets and magnetised outflows from the corona surrounding a thin disk with an increase in the magnetic diffusivity parameter. We have defined an equivalent time scale ($\tau_{eq}$) which takes into account the heating time scales due to viscosity, resistivity, magnetic reconnection and magneto-rotational instability turbulence such that the thin disk is formed if the cooling time scale ($\tau_{cool}$) is lower than this equivalent time scale ($\tau_{cool}/\tau_{eq}<1$). Using this condition, for the first time, we found that the thin disk exists when the initial ratio of plasma pressure to magnetic pressure (plasma beta) exceeds a range of $600--800$ for the gas obeying a polytropic equation of state accreting at $10^{-5}\ M_{\odot}/year$.
\end{abstract}


\maketitle


\section{\label{sec:Intro}Introduction}

Accretion onto black holes is an extremely powerful process capable of converting enormous amounts of gravitational potential energy into radiation. The most luminous objects in the universe, like quasars, are powered by accretion, with radiative efficiency of $0.1$ or more \cite{mahato13}. However, low luminosity Active Galactic Nuclei (AGN) like Seyferts are more abundant, where the densities of the accreting material are comparatively lower than quasars. In such accretion disks, the accretion flow can be broadly categorized into 2 types: A radiatively inefficient accretion flow (RIAF) where the radiation produced is unable to escape from the disk, causing flaring and thermal thickening \cite{ichimaru77,Stone99}; and an optically thick, geometrically thin disk that follows a blackbody spectrum \cite{shakura}. A sufficient amount of work has been done over the years to explain the formation of RIAF \cite{Igumen03,Machida04}. Rees et al. \cite{rees82} showed that the RIAF could transition into a thin disk if the cooling timescales become shorter than the viscous timescales and vice versa. The viscous stress in this thin disk can be treated as a perturbation to the local pressure \cite{shakura}.

An extensive amount of work has been done in the regime of thick-to-thin disk transition using high-resolution hydrodynamic (HD) and magnetohydrodynamic (MHD) simulations incorporating radiative processes \cite{Ohsuga09,sharma2013}. In the case of hot accretion flows, the radiative efficiency increases with the accretion rate and is comparable to a black body-dominated thin disk at high accretion rates \cite{Xie12}. As the radiative efficiency increases with accretion rate, the dominant radiation spectra change from synchrotron dominated to inverse-Compton \cite{Benjamin17}. However, the accretion rate for hot accretion flows decreases inward and is correctly explained by the adiabetic inflow-outflow solution (ADIOS), indicating that a thin disk may not exist very close to the black hole \cite{Yuan12}. Cooling-dominated accretion disks are thermally unstable without the presence of magnetic fields \cite{Xiao-fei15}. Advective cooling plays an important part in keeping optically thin disks thermally and viscously stable \cite{Abramowicz95}. Thermal conduction can give rise to the intermediate region between the hot corona and cool disk during the low-hard to high-soft transition \cite{Nakamura19}. Sadowski and Gaspari \cite{Sadowski17} performed GR-RMHD simulations and showed that the radiative output of accretion flow increases with accretion rate, and the state transition occurs earlier for hotter electrons. Das and Sharma, 2013 \cite{sharma2013} have carried out extensive HD simulations incorporating Bremsstrahlung cooling, where they explained the transients observed in black hole X-ray Binaries (BHXRBs). The magnetic field plays an important role in such a transition since there will be increased heating due to resistivity, reconnection and turbulent heating due to Magneto-Rotational Instability (MRI) while the magnetic pressure can support the disk during transition \cite{Machida06}. It is still unclear how these processes can affect the nature of geometrically thin accretion flow, although a significant amount of work has been done regarding the effect of magnetic fields on the accretion flow and jet launching \cite{balbus91,balbus98,hawley00,Hawley01,Ohsuga11,Mckinney12,Xue-Ning13,jacq24}.

Several attempts have been made to explain the observational properties of BHXRBs using numerical simulations. Begelman and Armitage \cite{begelman14} proposed that the increase in turbulent stress in disks threaded by a net magnetic field and the ability of thick (but not thin) disks to advect such a field radially is responsible for the hysteric cycle of black hole state transitions. Wu et al. \cite{wu16} carried out HD simulations coupled with Bremsstrahlung, synchrotron and Compton cooling to explain the observed transitional properties in BHXRBs. The time variability introduced by the quasi-periodic transition of gas from efficient cooling to inefficient cooling \cite{hogg17,hogg18} is responsible for spectro-temporal variabilities observed in accreting black holes. 3D global simulations of RIAF  were carried out by Dhang and Sharma \cite{sharma19}, 2019 to study the long-term evolution of non-radiative geometrically thick accretion flows and the dynamo process where they showed the effects of dynamical quenching of dynamo-$\alpha$ in global simulations of accretion flows. Recently, Dihingia et al. \cite{sharma23} performed General Relativistic MHD (GRMHD) simulations of geometrically thin accretion flows to study the decaying phase of outbursts in BHXRBs which led them to observe quasi-periodic oscillations (QPOs).

In this work, we have focused on the numerical evolution of the magnetised accretion flow in the parametric space to understand the effect of a magnetic field on the formation of a thin disk. In order to probe such a feature, we have varied the magnitude of the seed magnetic field along with the magnetic diffusivity and documented its effect on the overall nature of the flow. We have indeed found a threshold value for the seed magnetic field (52 gauss) at an accretion rate of $10^{-5}\ M_{\odot}/year$, which agrees with our simulations, beyond which the thin disk is no longer sustained. We have attributed this result to the overall increase in the heating rate due to the presence of magnetic fields. In order to account for this increase in heating rate, we have defined a new time scale ($\tau_{eq}$), which takes into account the time scales of all the heat sources. We assume that this time scale has to be higher than the cooling time scale for the existence of the thin disk. We have compared the threshold values of our model with our simulation to show that our model holds correct for such accretion flows.

The equations of the initial accretion torus are defined in Section \ref{sec:Equations}, the numerical setup is discussed in Section \ref{sec:setup}, the simulation runs are described in Section \ref{sec:Runs}, the results are presented in Section \ref{sec:Results}, the analysis of the flow structure is done in \ref{sec:Analysis} followed by our concluding remarks in Section \ref{sec:Conclusion}.

\section{\label{sec:Equations}Equations}

The initial accretion torus is assumed to be in a steady state with its azimuthal velocity ($V_{\phi}$) being specified by the Keplerian velocity. It is set using three parameters that specify the location and shape of the torus, namely the torus centre $(R_0)$ where the density is assumed to be maximum, the distortion parameter $(d)$, which determines the deviation of the torus cross-section from a circular profile and the scale height $(H)$ which determines the thickness of the torus. The accretion torus is set using the prescription by Papaloizou and Pringle \cite{pp84} with the initial equations being modified for the Paczyński–Wiita pseudo-Newtonian potential (Eq. \ref{eqn:potential}) to mimic general relativistic effects

\begin{equation}
    \label{eqn:potential}
    \phi=-\dfrac{GM}{r-R_g}
\end{equation}

where $G$ is the universal Gravitational constant, $M$ is the mass of the compact object, $r$ is the spherical polar radius and $R_g$ is the gravitational radius given by $2GM/c^2$. The pressure and density satisfy a polytropic equation of state (EOS) given by

\begin{equation}
    \label{eqn:eos}
    P=K\rho^{\gamma}
\end{equation}

where $P$ is the plasma pressure, $K$ is the polytropic constant, and $\gamma$ is the polytropic index, which specifies the internal thermodynamics involved. Thus, the modified equations from \cite{pp84} becomes

\begin{equation}
    \label{eqn:torus}
    \dfrac{P_t}{\rho_t}=\dfrac{GM}{(n+1)R_0}\left[\dfrac{R_0}{r-R_g}-\dfrac{R_0^4}{2\{(R_0-R_g)R\}^2}-\dfrac{1}{2d}\right]
\end{equation}

Here $R$ represents cylindrical polar radius and the subscript of $t$ is used to represent torus variables. When Eq. \ref{eqn:eos} is substituted in Eq. \ref{eqn:torus} for $r=R_0,\ \rho_t=\rho_0$, the polytropic constant becomes

\begin{equation}
    \label{eqn:poly-const}
    K=\dfrac{GM\rho_0^{(1-\gamma)}}{(n+1)R_0}\left[\dfrac{R_0}{R_0-R_g}-\dfrac{R_0^2}{2(R_0-R_g)^2}-\dfrac{1}{2d}\right]
\end{equation}

Thus, the expression for $K$ is used to specify $\rho_t$. The initial torus is fully specified by the following equations

\begin{equation}
    \label{eqn:rho}
    \rho_t= 
    \left[\frac{GM}{(n+1)R_0K}\left\{\frac{R_0}{r-R_g}-\frac{R_0^4}{2[(R_0-R_g)R]^2}-\frac{1}{2d}\right\}\right]^{\left(\frac{1}{\gamma-1}\right)}
\end{equation}

\begin{equation}
    \label{eqn:torus-eos}
    P_t=K\rho_t^{\gamma}
\end{equation}
\begin{equation}
    \label{eqn:init-eos}
    V_r=V_{\theta}=0 \\
\end{equation}
\begin{equation}
    \label{eqn:init-vphi}
    V_{\phi}=\sqrt{\dfrac{GM}{r}}\dfrac{r}{r-R_g}
\end{equation}

where $V_{r}$ and $V_{\theta}$ are the radial and meridional velocities respectively.

\section{\label{sec:setup}Numerical Setup}
The simulations are carried out in \href{https://plutocode.ph.unito.it/}{\texttt{PLUTO v4.4-patch2}} by Mignone et al. \cite{mig07}. It is a well-tested finite volume and finite difference solver which solves the set of HD and MHD conservation laws. The set of MHD equations which PLUTO solves is given as

\begin{widetext}
    \begin{equation}
    \label{eqn:mass}
    \dfrac{\partial\rho}{\partial t}+\nabla\cdot(\rho v)=0
\end{equation}

\begin{equation}
    \label{eqn:momentum}
    \dfrac{\partial m}{\partial t}+\nabla\cdot\left[mv-BB+\mathcal{I}\left(p+\frac{B^2}{2}\right)\right]^T=-\rho\nabla\Phi+\rho g
\end{equation}

\begin{equation}
    \label{eqn:induction}
    \dfrac{\partial B}{\partial t}+\nabla\times(cE)=0
\end{equation}

\begin{equation}
    \label{eqn:energy}
    \dfrac{\partial (E_t+\rho\Phi)}{\partial t}+\nabla\cdot\left[\left(\frac{\rho v^2}{2}+\rho e+p+\rho\Phi\right)v+cE\times B\right]=m\cdot g
\end{equation}
\end{widetext}

Eq. \ref{eqn:mass}, \ref{eqn:momentum}, \ref{eqn:induction} and \ref{eqn:energy} describe the time evolution of mass, momentum, magnetic field and energy, respectively. $\rho$, $v$ and $p$ represent the density, velocity and pressure primitive variables, respectively, while $m$, $B$, $E$, $\phi$, $E_t$ and $e$ represents the momentum, magnetic field strength, electric field strength, body-force potential, total energy density and internal energy density respectively.\\
\\
The simulations are performed in 2.5D spherical axis-symmetric coordinates using the \texttt{HLLC} solver. The \texttt{HLLC} solver has been preferred over other solvers because of its ability to restore the contact discontinuity while being computationally efficient. Since we have focused on non-relativistic MHD simulations, we defined the inner boundary of the radial computational domain at $r=5R_g$ to avoid relativistic velocities and Alfven speeds. The outer radial boundary is kept fixed at $400R_g$. The beginning of the computational domain in the $\theta$ direction was kept at $10^{-8}rad$ to avoid singularities due to $1/r\sin(\theta)$ terms. A logarithmically spaced gird is used in the radial direction in order to capture finer details near the central object. The mass of the black hole is fixed at $10^6\ M_{\odot}$. All quantities are scaled by typical values of density, velocity and length to avoid extremely large or small values during computation. Density is scaled using the density specified at the torus centre $(\rho_0)$, length is taken in units of $R_g$ and velocity is scaled using the Keplerian velocity at $R_g$

\begin{equation}
    \label{eqn:kep-v}
    V_{K,R_g}=\sqrt{\dfrac{GM_{\odot}}{R_g}}
\end{equation}
\\
Viscosity and resistivity terms are treated explicitly for runs where $\nabla\cdot B=0$ condition is preserved using \texttt{CONSTRAINED TRANSPORT} \cite{CT04}. A Van-Leer limiter is used when reconstruction is set to linear and \texttt{SHOCK\_FLATTENING} is set to \texttt{MULTID} in the \texttt{definitions.h} file. The multidimensional shock flattening strategy enables switching to the diffusive \texttt{HLL} \cite{hll07} solver and uses a mind limiter in the presence of strong shocks \cite{MM86}. In order to prevent negative pressures and negative densities due to rarefaction waves, a floor value of $10^{-10}$ was set for both quantities. In order to prevent negative values of total energy, \texttt{FAILSAFE} is enabled in \texttt{definitions.h}, which allows us to retry the step with flat reconstruction and \texttt{HLL} solver. A scaler tracer is set to track the torus material and to specify the viscosity and resistivity in only the torus material. We have used resolutions of $(\mathcal{R\times\theta})=(512\times512)$, $(256\times256)$ and $(128\times128)$ depending on the value of time step. All runs are listed in Table \ref{tab:table:runs}. The torus has been allowed to evolve till 100 days, by then the accretion rate reaches a steady state (Fig. \ref{fig:hd_v_mhd-accretion-rate}).

\subsection{\label{sub-sec:init}Initial Conditions}

The initial equations after scaling are given by

\begin{equation}
    \label{eqn:scale-poly-index}
    K=\dfrac{1}{(n+1)R_0}\left[\dfrac{R_0}{R_0-1}-\dfrac{R_0^2}{2(R_0-1)^2}-\dfrac{1}{2d}\right]
\end{equation}

\begin{equation}
    \label{eqn:scale-rho}
    \rho_t=\left[\dfrac{1}{(n+1)R_0K}\left\{\dfrac{R_0}{r-1}-\dfrac{R_0^4}{2[(R_0-1)R]^2}-\dfrac{1}{2d}\right\}\right]^{\frac{1}{\gamma-1}}
\end{equation}

\begin{equation}
    \label{eqn:scale-prs}
    P_t=K\rho_t^{\gamma}
\end{equation}

\begin{equation}
    \label{eqn:scale-vphi}
    v_{\phi}=\dfrac{\sqrt{r}}{r-1}
\end{equation}

\begin{equation}
    \label{eqn:scale-potential}
    \phi=-\dfrac{1}{r-1}
\end{equation}

with $\gamma=5/3$. The torus was set up at $R_0=100R_g$ with $d=1.125$ and a scale height of $0.1$. The initial torus is assumed to be surrounded by a static ambient atmosphere of constant density $\rho_a=10^{-4}$ and pressure specified by

\begin{equation}
    \label{eqn:amb_prs}
    \rho_a=P_aH
\end{equation}

Where $a$ denotes ambient medium variables.

\subsection{\label{sub-sec:boundary}Boundary Conditions}

The inner radial boundary condition is set to inflow by setting the radial velocity as $MIN(v_r,0)$, and outflow is set at the outer radial boundary condition by setting the radial velocity as $MIN(0,v_r)$ at the boundary. Unlike Das and Sharma, 2013 \cite{sharma2013}, viscosity was prescribed in the inner boundary to prevent discontinuities. All other quantities are prescribed from the computational domain. Axis-symmetry has been set for $\theta_{start}$ and $\theta_{end}$. The azimuthal boundary conditions are specified as zero gradients in either direction.

\subsection{\label{sub-sec:viscosity}Viscosity}

The viscosity in our work has been prescribed using the viscous stress tensor. In this setup, only the $\tau_{r\phi}=\tau_{\phi r}$ of the viscous stress tensor is important since this is the component responsible for angular momentum transfer and hence accretion. Moreover, this is the most dominant component of magnetic stress in the magnetic stress tensor, which we have defined in Sec. \ref{sub-sec:resistivity} \cite{hawley95}. This component of the viscous stress is given by

\begin{equation}
    \label{eqn:stress-rphi}
    \tau_{r\phi}=\eta\left(\dfrac{\partial v_{\phi}}{\partial r}-\dfrac{v_{\phi}}{r}+\dfrac{1}{r\sin{\theta}}\dfrac{\partial v_r}{\partial \phi}\right)
\end{equation}

Since we have considered an axis-symmetric setup $\partial /\partial\phi=0$, this component gets modified to

\begin{equation}
    \label{eqn:stress-rphi-mod}
    \tau_{r\phi}=\eta\left(\dfrac{\partial v_{\phi}}{\partial r}-\dfrac{v_{\phi}}{r}\right)
\end{equation}

$\eta$ is the dynamic viscosity coefficient. $\eta$ is related to the kinematic viscosity coefficient as $\eta=\rho\nu$.

The Shakura-Sunyaev $\alpha$ thin disk model specifies $\nu$ as $\nu=\alpha c_s H$ \cite{shakura}, where $c_s$ is the local sound speed. For a thin accretion disk, this is assumed to be constant, which is not true in general. Alternatively, the disk height can be expressed as $H(r)\approx c_s/\Omega(r)$ where $\Omega(r)$ is the angular velocity at $r$. This means, $\nu=\alpha H^2 \Omega$ which is similar to the $\beta$ prescription provided in Eq. 1 of \cite{duschl00}. For a Keplerian orbit, $\Omega \propto r^{-3/2}$  and assuming a constant aspect ratio, the kinematic viscosity parameter $\nu$ varies as $r^{1/2}$ \cite{sharma2013}. Thus, in this work, we have considered the following form of kinematic viscosity:

\begin{equation}
    \label{eqn:nu}
    \nu_v=0.01\alpha_v\Omega_0R_0^2\sqrt{\dfrac{r}{R_0}}
\end{equation}
\\
where a subscript of $0$ indicates values at the torus center and $\alpha_v$ is a dimensionless number $<1$ This means that $\nu_v\propto r^{1/2} \implies \eta\propto \rho r^{1/2}$ throughout the computational domain and is not constant.\\
\\
The $\tau_{r\phi}$ and $\tau_{\phi r}$ components have been specified by modifying the \texttt{viscous\_flux.c} file.~All other components in the file are set to 0. The bulk viscosity component $(\eta_2)$ is set as $0$, and the shear viscosity component $(\eta_1)$ was set by multiplying $\rho$ to Eq \ref{eqn:nu}. The scalar tracer was multiplied to $\eta_1$ so that the viscosity is only specified for the torus material.

\subsection{\label{sub-sec:cooling}Cooling}

Radiative transfer has not been considered in our work since treating such processes is computationally intensive. Optically thin bremsstrahlung cooling has been used instead \cite{sharma2013}. The cooling function is incorporated in the energy equation as $n_en_i\Lambda(T)\approx n^2\Lambda((T)$. As prescribed by Das and Sharma, 2013 \cite{sharma2013}, $\Lambda(T)$ is specified using the function in equation 12 of \cite{cool10}. In this work, we assume that the disk transition occurs when the sum of the viscous heating rate, the resistive heating rate, the magnetic reconnection heating rate and the MRI heating rate becomes equal to the cooling rate. In this case, the thick RIAF separates into a cool, thin disk and a hot coronal phase \cite{malzac2006,coppi99}. In \textsc{PLUTO}, tabulated cooling has been specified, which uses piece-wise polynomials to interpolate the cooling function at specific temperatures. The cool table was constructed by supplying a uniformly spaced array of temperatures ranging from $10^3K$ to $10^{13}K$ to the $\Lambda(T)$ function.

\subsection{\label{sub-sec:Mag-field}Magnetic Field}

The magnetic field has been specified as a local poloidal field present only in the torus, which is exponentially damped at the torus boundary, which prevents any initial discontinuities in $\vec{B}$ and therefore preserves $\nabla\cdot B=0$. The field in all the cases has the form $\vec{B}(R)=B_0(R)\hat{z}$ where $B_0$ is a spatially varying magnitude. $B_0$ is of the form $\mu f(R)$, with $f(R)=1/R$. The value of $\mu$ has been varied to change the magnitude of the seed magnetic field. 

\subsection{\label{sub-sec:resistivity}Resistivity}

A prescription similar to the viscosity is used for resistivity as well. The resistivity $(\eta_m)$ is given by

\begin{equation}
    \label{eqn:eta}
    \eta_m=4\pi\nu_m
\end{equation}

Where $\nu_m$ is the magnetic diffusivity, which, according to the $\alpha$, is given as

\begin{equation}
    \label{eqn:nu_m}
    \nu_m=\dfrac{3}{2}\alpha_m\dfrac{\nu_v}{\alpha_v}
\end{equation}

where $\alpha_m$ is a dimensionless number $<1$.

\section{\label{sec:Runs}Runs}

\begin{table}
    \centering
    \begin{tabular}{ccccccc}
        \hline
        &  &  &  &  &  &  \\
        Run & Resolution & $\mu$ & $\beta$ & $\alpha_m$ & $\alpha_v$ & $\rho(m_p/cc)$ \\
        &  &  &  &  &  &  \\
        \hline
        &  &  &  &  &  &  \\
        H.09 & $512^2$ & -- & -- & -- & 0.01 & $10^{9}$ \\
        H.10 & $512^2$ & -- & -- & -- & 0.01 & $10^{10}$ \\
        H.11 & $512^2$ & -- & -- & -- & 0.01 & $10^{11}$ \\
        B.01 & $512^2$ & 0.01 & >5000 & 0.01 & 0.01 & $10^{10}$ \\
        B.1 & $128^2$ & 0.1 & >5000 & 0.01 & 0.01 & $10^{10}$ \\
        AV.1.01 & $256^2$ & 0.01 & >5000 & 0.01 & 0.1 & $10^{10}$ \\
        AV.1.1 & $128^2$ & 0.1 & >5000 & 0.01 & 0.1 & $10^{10}$ \\
        AM.09.01 & $128^2$ & 0.1 & >5000 & 0.1 & 0.01 & $10^{9}$ \\
        AM.10.01 & $128^2$ & 0.1 & >5000 & 0.1 & 0.01 & $10^{10}$ \\
        AM.11.01 & $128^2$ & 0.1 & >5000 & 0.1 & 0.01 & $10^{11}$ \\
        AM.10.1 & $128^2$ & 0.1 & >5000 & 0.1 & 0.1 & $10^{10}$ \\
        AM.11.1 & $128^2$ & 0.1 & >5000 & 0.1 & 0.1 & $10^{11}$ \\
        D.09.01 & $512^2$ & 0.01 & >5000 & 0.01 & 0.01 & $10^{9}$ \\
        D.09.1 & $128^2$ & 0.1 & >5000 & 0.01 & 0.01 & $10^{9}$ \\
        D.10.01 & $512^2$ & 0.01 & >5000 & 0.01 & 0.01 & $10^{10}$ \\
        D.10.1 & $128^2$ & 0.1 & >5000 & 0.01 & 0.01 & $10^{10}$ \\
        D.11.01 & $256^2$ & 0.01 & >5000 & 0.01 & 0.01 & $10^{11}$ \\
        B.010.2 & $128^2$ & 0.2 & 1394 & 0.01 & 0.01 & $10^{10}$ \\
        B.010.3 & $128^2$ & 0.3 & 622 & 0.01 & 0.01 & $10^{10}$ \\
        B.025.4 & $128^2$ & 0.4 & 348 & 0.025 & 0.01 & $10^{10}$ \\
        B.025.5 & $128^2$ & 0.5 & 223 & 0.025 & 0.01 & $10^{10}$ \\
        B.050.6 & $128^2$ & 0.6 & 154 & 0.05 & 0.01 & $10^{10}$ \\
        B.050.7 & $128^2$ & 0.7 & 114 & 0.05 & 0.01 & $10^{10}$ \\
        B.075.8 & $128^2$ & 0.8 & 87 & 0.075 & 0.01 & $10^{10}$ \\
        B.075.9 & $128^2$ & 0.9 & 69 & 0.075 & 0.01 & $10^{10}$ \\
        &  &  &  &  &  &  \\
        \hline
    \end{tabular}
    \caption{List and details of runs. \texttt{H}, \texttt{AV}, \texttt{AM}, \texttt{D} and \texttt{B} represents HD and runs with variation in $\alpha_v$, $\alpha_m$, $\rho$ and $\mu$ respectively. $\beta$ represents the initial plasma beta value at a cylindrical radius of $R=100R_g$.}
    \label{tab:table:runs}
\end{table}

The list of runs can be found in Table \ref{tab:table:runs}. The magnetic field at the edges has been smoothed out to preserve $\nabla\cdot B=0$. Runs \texttt{H.09}, \texttt{H.10} and \texttt{H.11} are purely hydrodynamic simulations. Runs \texttt{B.01} and \texttt{B.1} help us to identify the strength of the seed magnetic field, which starts affecting the flow. Runs with the \texttt{AV} prefix focus on the changes in the flow characteristics due to $\alpha_v=0.1$. We have investigated the effect of magnetic diffusivity on the optically thick-geometrically thin flow for different accretion rates for runs with prefix \texttt{AM}. The accretion rate dependence of the flow structure in the presence of the magnetic field is investigated in runs with prefix \texttt{D}. Finally, we have investigated the effect of the magnetic field on the disk transition from runs \texttt{B.01.2} to runs \texttt{B.075.9} where we have varied the magnetic field strength along with $\alpha_m$ to find trends in the upper limit of plasma beta $\beta$ (ratio of plasma pressure to magnetic pressure)for which such a geometrically thin disk can be sustained. In these runs, the inner radial computational boundary has been limited to $r=15R_g$ in order to prevent relativistic Alfven speeds, which make the computation extremely slow. The accretion rate has not been documented for these runs since the inner boundary is far from ISCO, and such an inner radius cutoff does not affect the disk transition, which is our primary concern in this paper. However, coronal jet launching is not observed for such a choice of the inner boundary, and these will be addressed in future works.

\section{\label{sec:Results}Results}

Accretion rates are normalised in units of $\rho_0R^2V_0$ and are calculated as $\dot{M}=\langle r\int_S\rho\vec{V}\cdot \vec{dS}\rangle_r$ after discretizing the equation according to the grid. The accretion rate described here is the radial distance weighted average accretion rate from $5R_g$ to $10R_g$, and $dS$ are the discretized surfaces of the sphere with radius $r$. The disk height is given as $H\approx c_s/\Omega$ where $c_s$ is the sound speed and $\Omega$ is the angular velocity. The sound speed can be approximated as $c_s\approx\sqrt{\gamma P/\rho}$ for an ideal EOS. With the angular velocity given as $\Omega\approx\sqrt{GM/R^3}$, we have the aspect ratio as

\begin{equation}
    \label{Aspect ratio}
    \dfrac{H}{R}=\sqrt{\dfrac{\gamma PR}{\rho}}
\end{equation}

The magnetic field is scaled by $\sqrt{4\pi\rho_0V_0^2}$ which is proportional to $\sqrt{\rho_0}$ and is thus different for runs with different $\rho_0$. This corresponds to $5.05\times 10^{3}$, $1.60\times 10^{4}$ and $5.05\times 10^4$ gauss for $10^9$, $10^{10}$ and $10^{11}\ m_p/cc$ respectively.

\subsection{\label{Res:HD}Hydrodynamic}

\begin{figure*}
    \centering
    \includegraphics[width=1.0\linewidth]{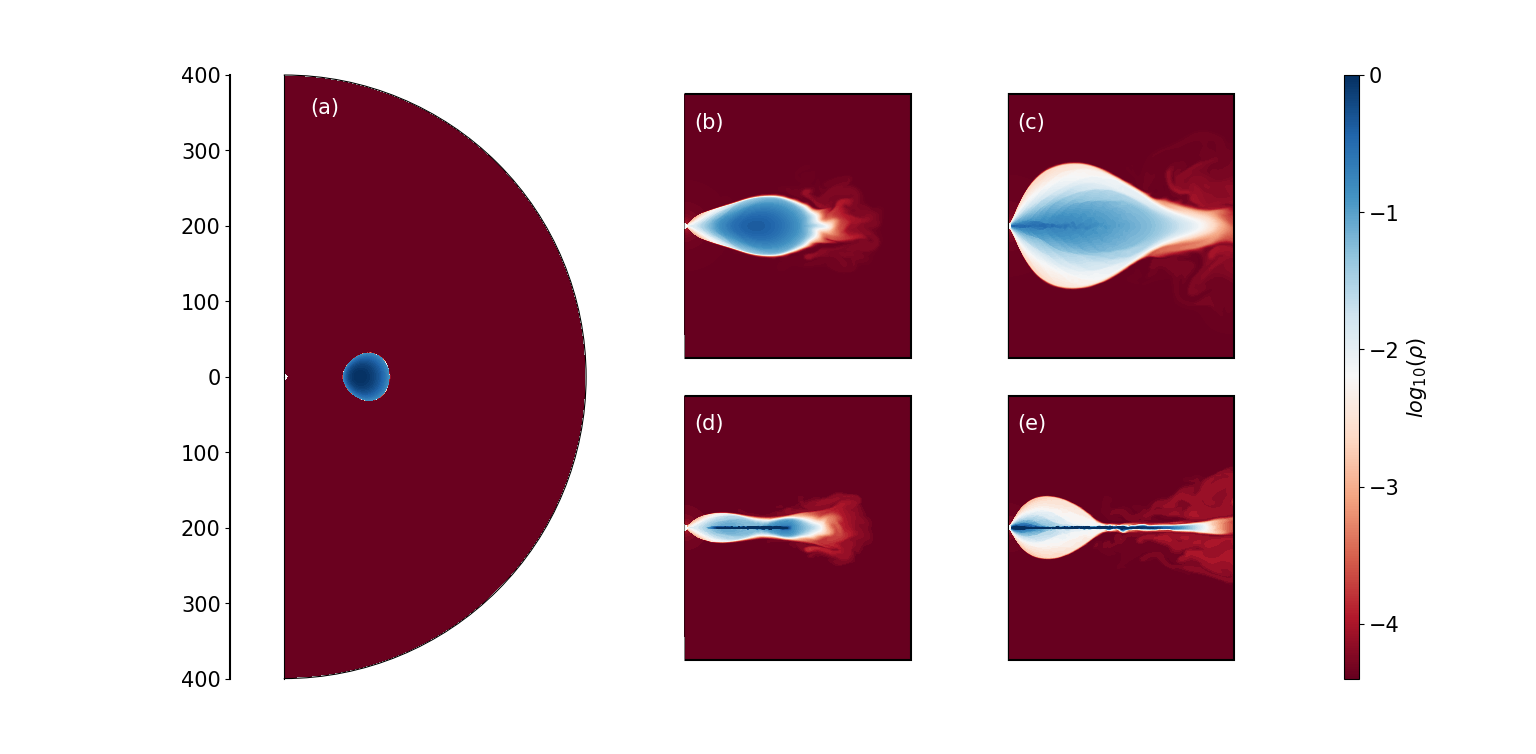}
    \caption{Density contour plots demonstrating the evolution of the torus. Figure (a) is the initial torus. Figures (b) and (c) describe the flow for $\rho_0=10^9m_p/cc$ after $23$ days and $100$ days, respectively. Figures (d) and (e) describe the flow for $\rho_0=10^10m_p/cc$ after $23$ and $100$ days, respectively. We see the formation of the cold thin disk in panel (d) and (e) where the cooling rate has exceeded the viscous heating rate. The y range is in units of $R_g$.}
    \label{fig:HD_basic_rho}
\end{figure*}

\begin{figure*}
    \centering
    \includegraphics[width=1.0\linewidth]{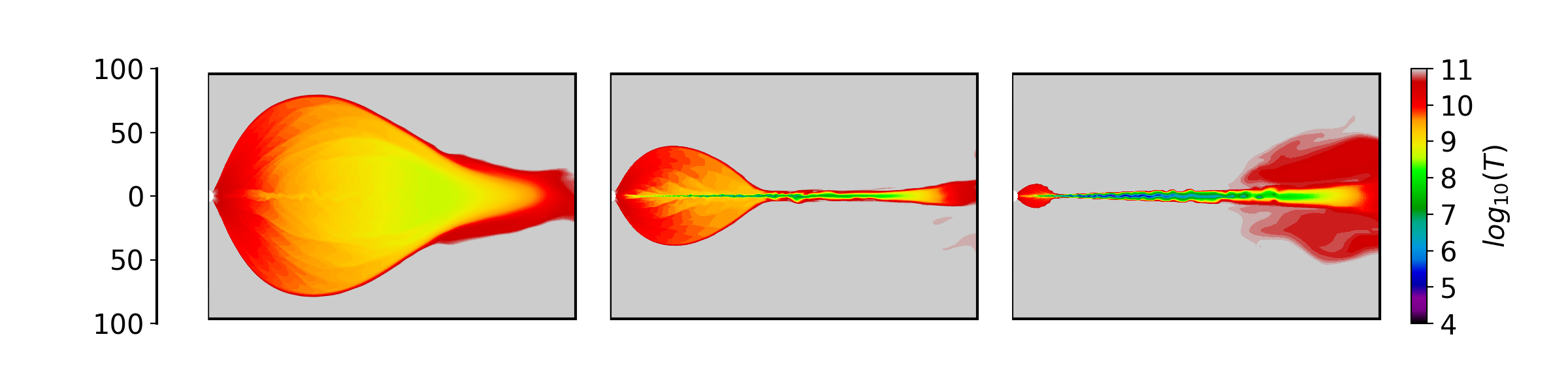}
    \caption{Temperature profile for runs \texttt{H.9}, \texttt{H.10} and \texttt{H.11} after $t=100$ days. The disk temperature decreases with a higher accretion rate, with the lowest temperature for run \texttt{H.11} being half an order of magnitude lower than that of run \texttt{H.10}. The corona for run \texttt{H.10} extends till $100R_g$ and till $25R_g$ for run \texttt{H.11} along the $x$ direction. The y range is in units of $R_g$.}
    \label{fig:HD-temp}
\end{figure*}

In this section, we present the purely hydrodynamic case. Run \texttt{H.09} describes the RIAF where the cooling rate is lower than the viscous heating rate (Fig. \ref{fig:HD_basic_rho}). Run \texttt{H.10} marks the onset of the disk transition (Fig. \ref{fig:HD_basic_rho}) where the RIAF separates out into an optically thick geometrically thin cool phase and a hot corona extending up to $30R_g$ in the vertical direction. This transition is triggered by the higher density of the initial torus and, hence, a higher rate of cooling. The accretion flow evolves from a peanut-like shape into a thin disk, as shown in \cite{sharma2013}. A similar phenomenon is observed in run \texttt{H.11}, where the transition occurs faster. This is because the cooling rate is higher as it varies as $n^2$ (Sec. \ref{sub-sec:cooling}). In this case, the coronal region is found to be smaller (extending up to $7R_g$ in the vertical direction) and agrees with observational data where the evidence of the coronal size decreases with an increase in mass accretion rate of supermassive black holes \cite{coppi99,malzac2006,belmont08}. The cold-thin disk average temperature decreases with the accretion rate (Fig. \ref{fig:HD-temp}). The simulations return a temperature range of $10^4-10^5\ K$ for an accretion rate of $10^{-5}\ M_{\odot}/year$ and $3.98\times 10^{3}-10^5\ K$ for an accretion rate of $8\times10^{-4}\ M_{\odot}/year$.

\subsection{\label{Res:Magnetic}Magnetic field}

\begin{figure*}
    \centering
    \includegraphics[width=1.0\linewidth]{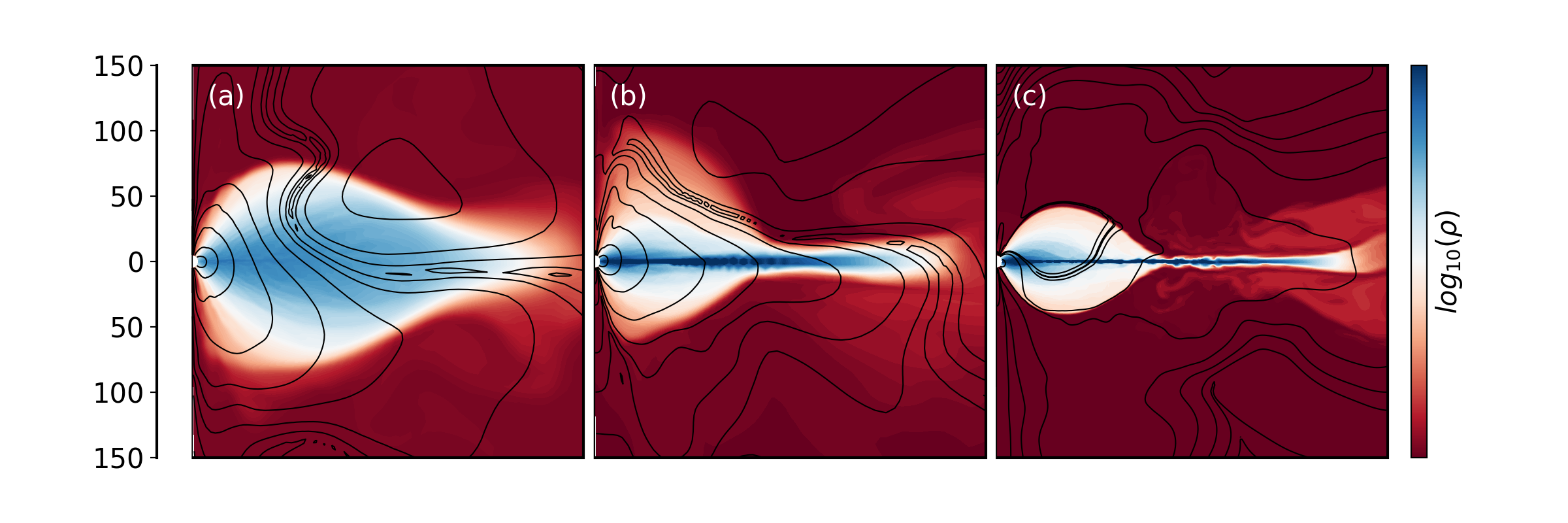}
    \caption{Density and magnetic field contours for runs \texttt{D.09.1}, \texttt{D.10.1} and \texttt{D.10.01} respectively after $100$. The magnetic contours are primarily dominated by the toroidal component. The hot flow gets skewed in the runs with $\mu=0.1$, and the magnetic contours conform to the structure of the hot flow, indicating that the toroidal component of the magnetic field is responsible for this behaviour. The y range is in units of $R_g$.}
    \label{fig:Mag-rho}
\end{figure*}

\begin{figure*}
    \centering
    \includegraphics[width=1.0\linewidth]{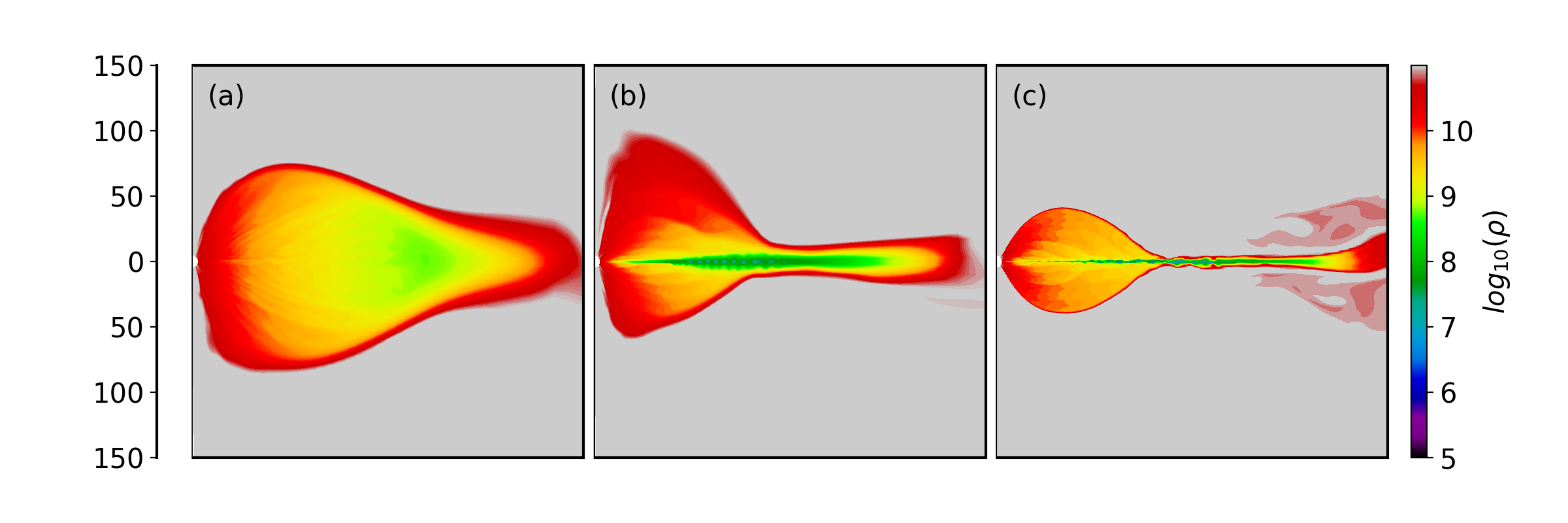}
    \caption{Temperature contours after $t=100$ days for runs \texttt{D.09.1}, \texttt{D.10.1} and \texttt{D.10.01} respectively. The temperature of the thin disk is of the order of $10^7-10^8K$ for runs with $\mu=0.1$ while it is of the order of $10^6-10^7K$ for $\mu=0.01$. The y range is in units of $R_g$.}
    \label{fig:mag-temp}
\end{figure*}

\begin{figure*}
    \centering
    \includegraphics[width=1.0\linewidth]{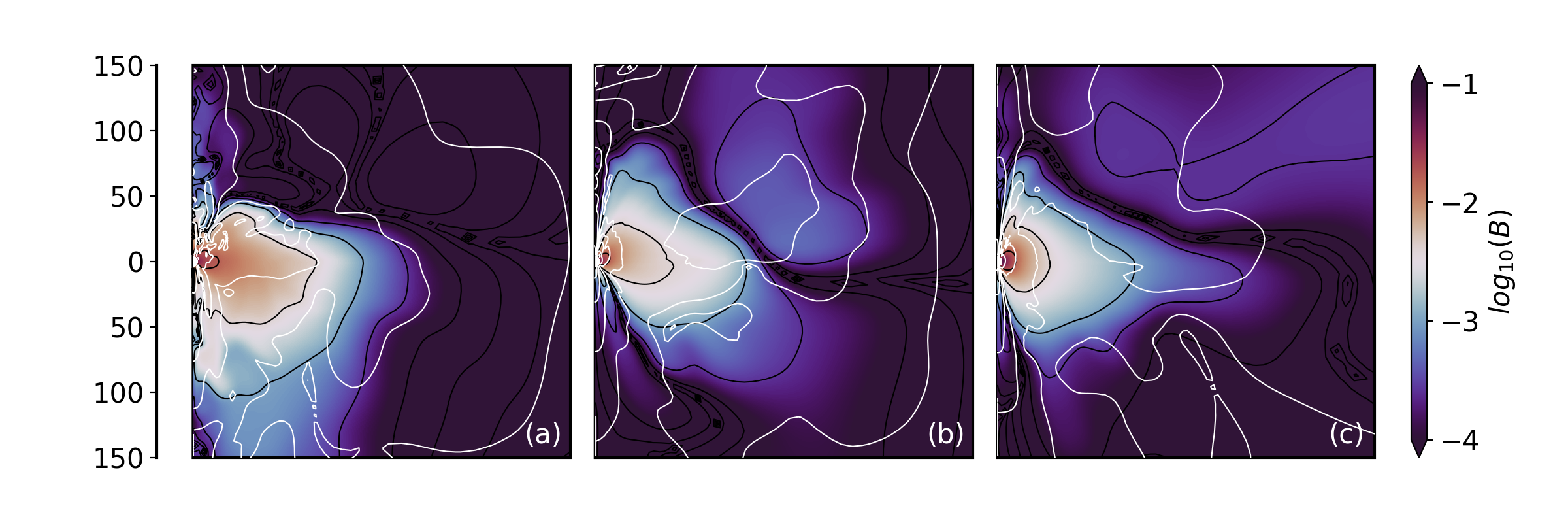}
    \caption{Evolution of magnetic field with for \texttt{B.1} after at $t=23,\ 50$ and $100$ days respectively. The black contour lines represent the toroidal component of the magnetic field, while the white contour lines represent the magnetic field strength in the $r-\theta$ plane. The toroidal contours are in the shape of the accretion flow, while the poloidal fields get advected inward and become concentrated along the axis. The y range is in units of $R_g$.}
    \label{fig:Magf}
\end{figure*}

\begin{figure*}
    \centering
    \includegraphics[width=0.49\linewidth]{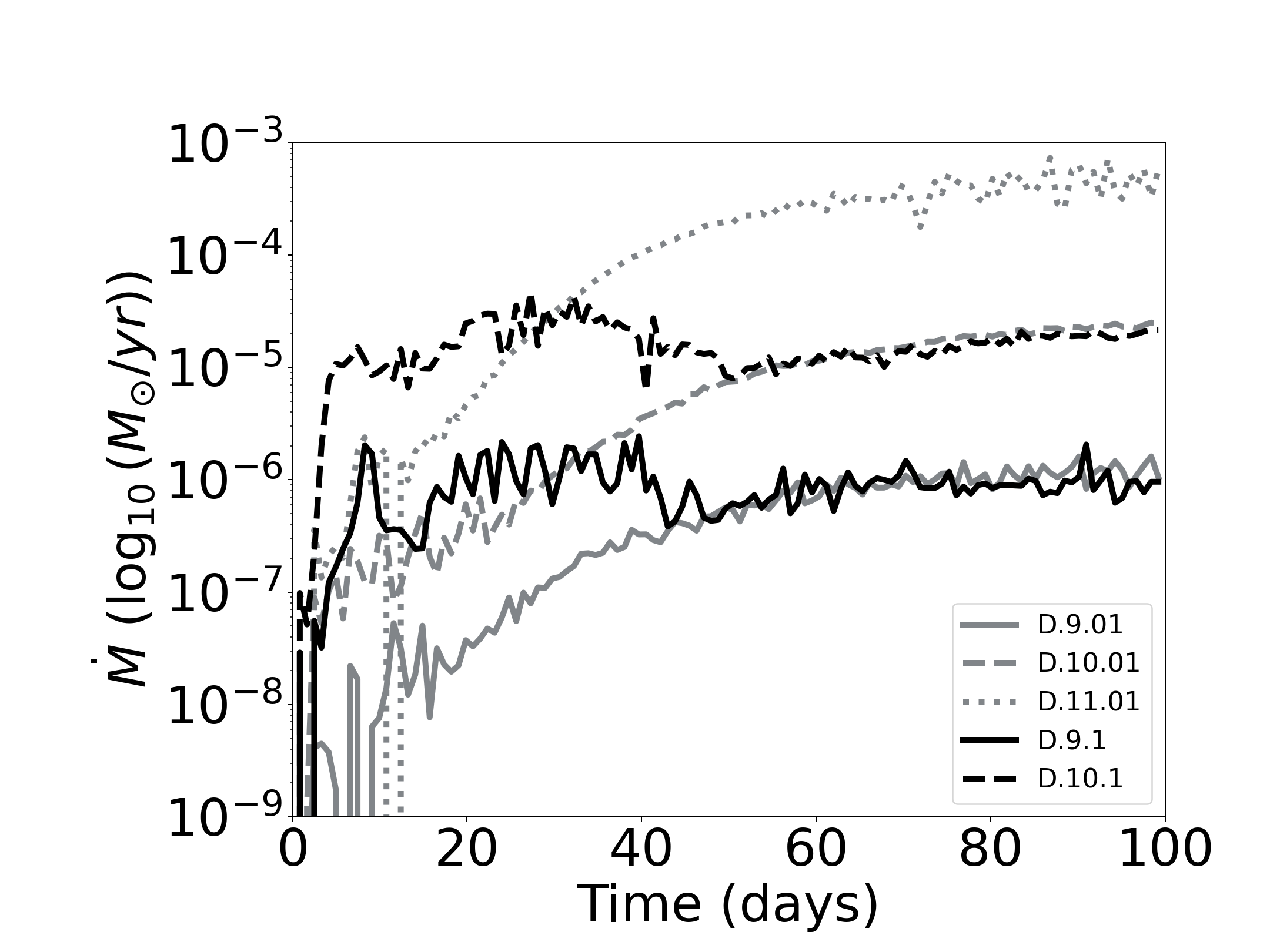}
    \includegraphics[width=0.49\linewidth]{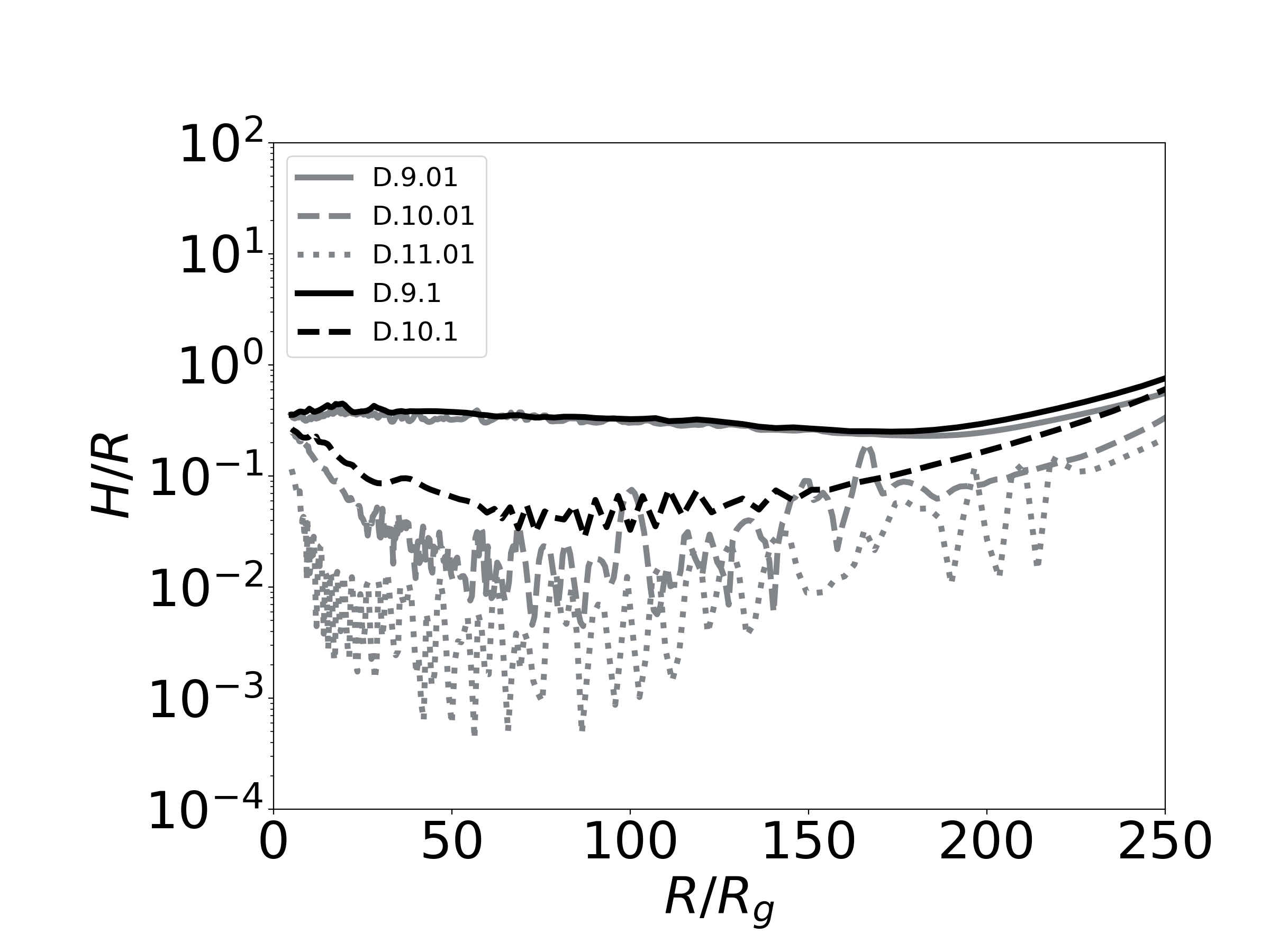}
    \caption{Left panel: Accretion rate vs time for comparatively higher initialisation magnetic field strength runs. There is an initial jump in the accretion rate for higher $\mu$ values due to the production of the MRI. Right panel: Midplane aspect ratio plots show that the aspect ratio has started increasing with an increase in magnetic field strength (indicated by a dashed black line).}
    \label{fig:rho-accretion-rate}
\end{figure*}

In the runs with an initial seed magnetic field, a value 0f $\mu=0.1$ acts as a threshold beyond which the magnetic pressure becomes comparable to the thermal pressure. In terms of plasma beta ($\beta$), it is of the order of $\beta=10^2$. Run \texttt{B1.01} (Fig. \ref{fig:Mag-rho}(c)) is very similar to \texttt{H.10} boasting a similar accretion rate of $10^{-5}\ M_{\odot}/year$ and a similar average aspect ratio of $0.01$ (Fig. \ref{fig:hd_v_mhd-accretion-rate}). The magnetic field in the $r-\theta$ plane gets diffused as it gets advected inward and forms a small concentration near the axis with a peak value of $4.61$ gauss at the beginning of the computational boundary and decreases exponentially as we move radially outwards. A strong toroidal field of magnitude $\approx 1500$ gauss develops at $10R_g$ and less due to the orbit of the plasma (Fig. \ref{fig:Magf}). For run \texttt{B.1} (Fig. \ref{fig:Mag-rho}(b)), the corona gets skewed due to magnetic pressure exerted by the toroidal fields(Fig. \ref{fig:Mag-rho}) with the skewness varying as a function of time on either side of the midplane. The coronal thickness varies from $50R_g$ to $100R_g$ due to variation of the toroidal field as a function of time. Low-density magnetized outflows have been observed originating from the skewed coronal material within a timescale of hours. These outflows are a result of the magnetic field being advected and concentrated near the axis along with the production of toroidal fields (Fig. \ref{fig:Magf}), giving rise to twisted long-range poloidal fields of the order of $10$ gauss or more extending beyond $250R_g$. This run shows an increase in the temperature of the thin disk from $10^5-10^6\ K$ to $10^6-10^7\ K$ (Fig. \ref{fig:mag-temp}) and average aspect ratio from $0.01$ to $0.08$ (Fig. \ref{fig:rho-accretion-rate}) when compared to run \texttt{B1.01} which may be due to an increase in magnetic pressure which drives faster Alfven waves increasing the rate of reconnection heating in the process.

\subsection{\label{Res:Denisty}Density}

\begin{figure*}
    \centering
    \includegraphics[width=0.49\linewidth]{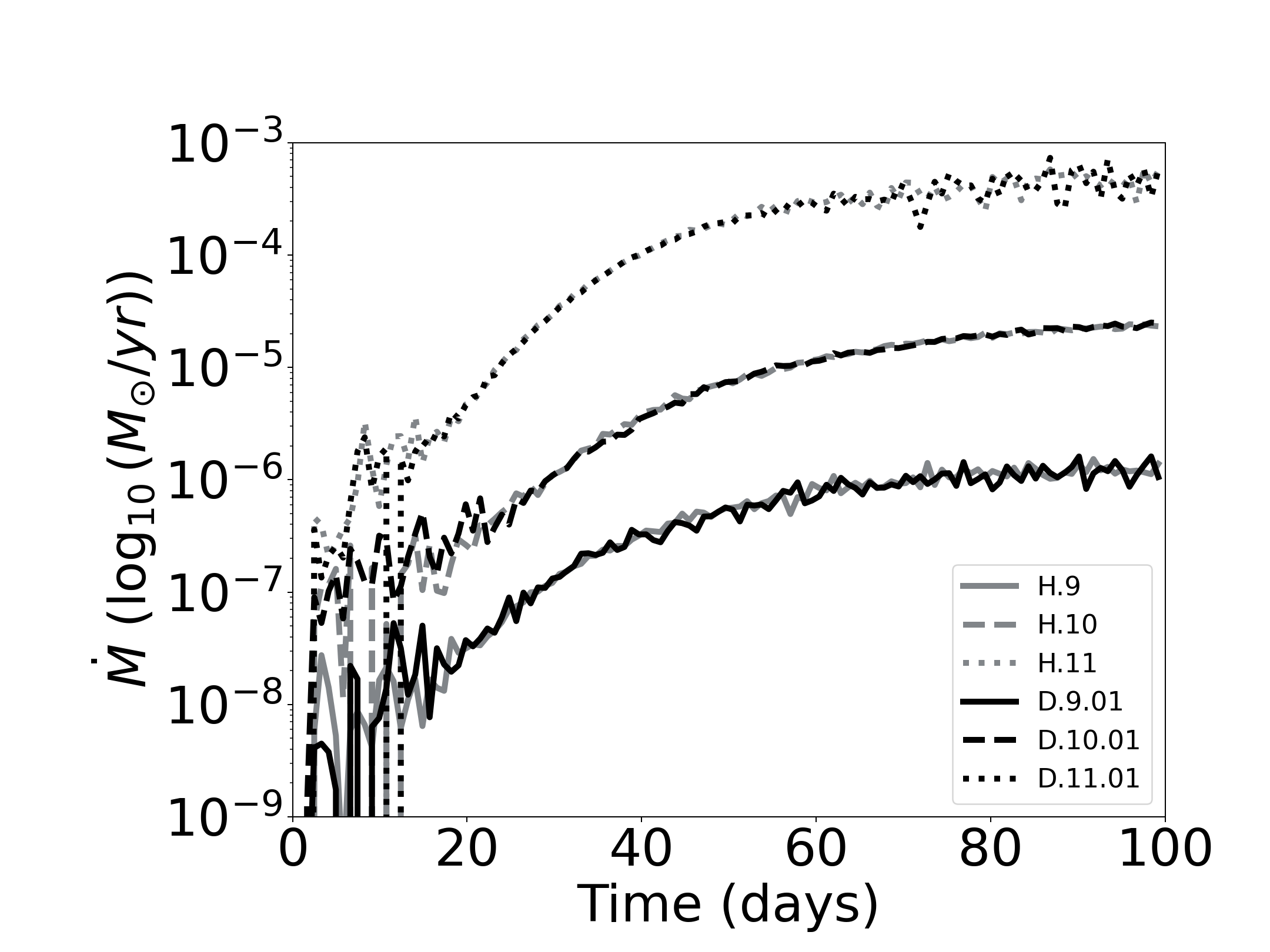}
    \includegraphics[width=0.49\linewidth]{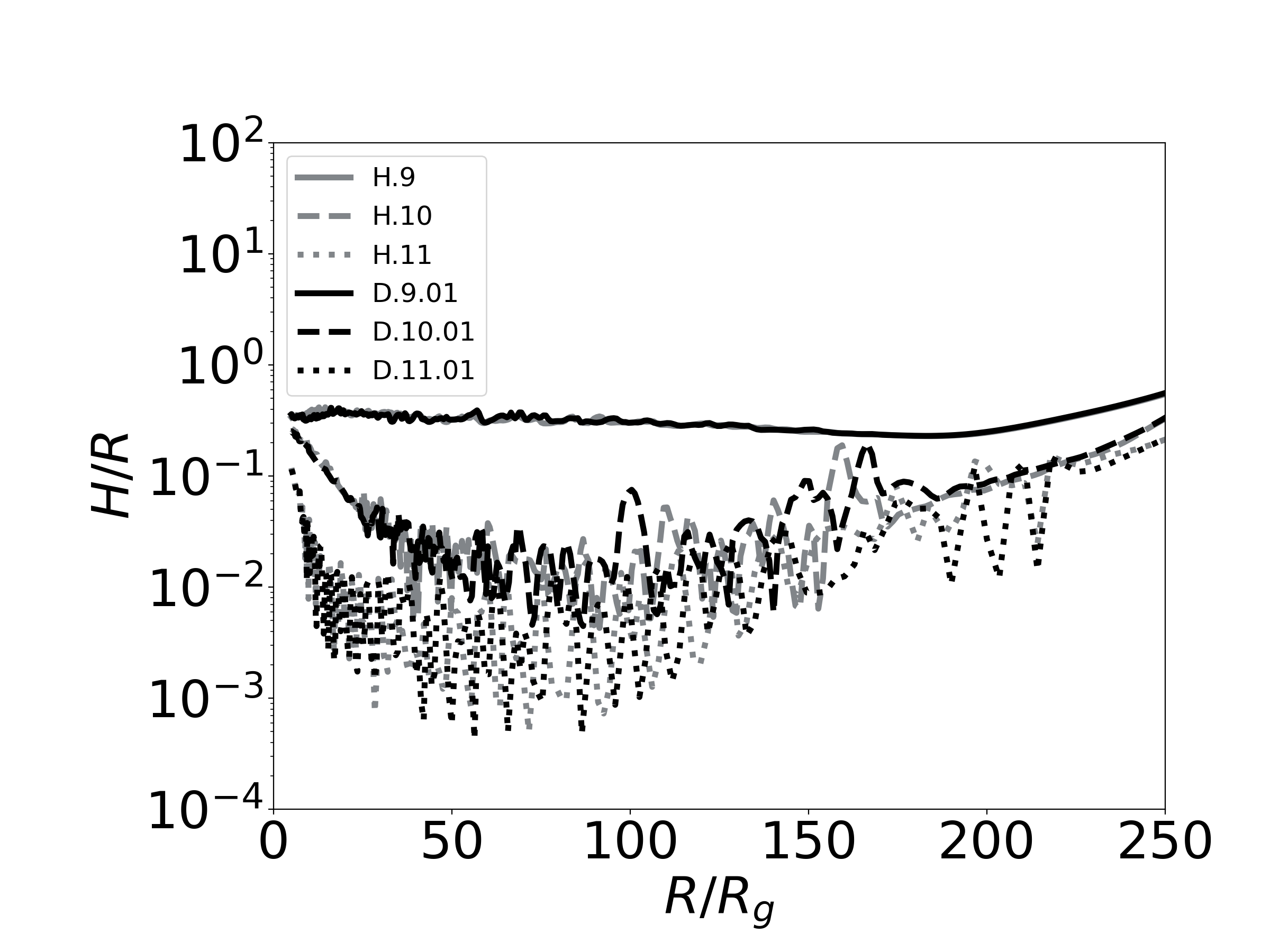}
    \caption{Left panel: Accretion rate vs time comparison of HD and MHD. At low magnetic field strengths, the accretion rate is not affected by the presence of magnetic fields. Right panel: Aspect ratio plots HD and MHD runs for low magnetic field strength. Both the runs report similar aspect ratio profiles.}
    \label{fig:hd_v_mhd-accretion-rate}
\end{figure*}

The density serves as a proxy for the accretion rate since it is directly proportional to the density (Section. \ref{sec:Results}). Higher accretion rates imply higher cooling rates since the cooling function is proportional to $n^2$ (Sec. \ref{sub-sec:cooling}). Thus, Figure \ref{fig:HD-temp} indicates a thin disk temperature range of $10^4-10^5\ K$ corresponding to an accretion rate of $5\times 10^{-4}\ M_{\odot}/year$ (Figure \ref{fig:hd_v_mhd-accretion-rate}) and $10^5-10^6\ K$ corresponding to an accretion rate of $10^{-5}\ M_{\odot}/year$. The effect of the hot corona decreases with a higher accretion rate and is similar to the hydrodynamic case for a $\mu$ value of $\leq0.01$. The magnetic field in each case is advected where it forms a poloidal field near the axis and toroidal fields with $\beta\approx1000$ up to $100R_g$. Both the accretion rate and the aspect ratio are similar to the hydrodynamic case (Fig. \ref{fig:hd_v_mhd-accretion-rate}), indicating that a $\mu$ value of $0.01$ is insufficient in producing a magnetic pressure comparable to the thermal pressure.

\subsection{\label{Res:alpha_m}Magnetic diffusivity ($\alpha_m$)}

\begin{figure*}
    \centering
    \includegraphics[width=1.0\linewidth]{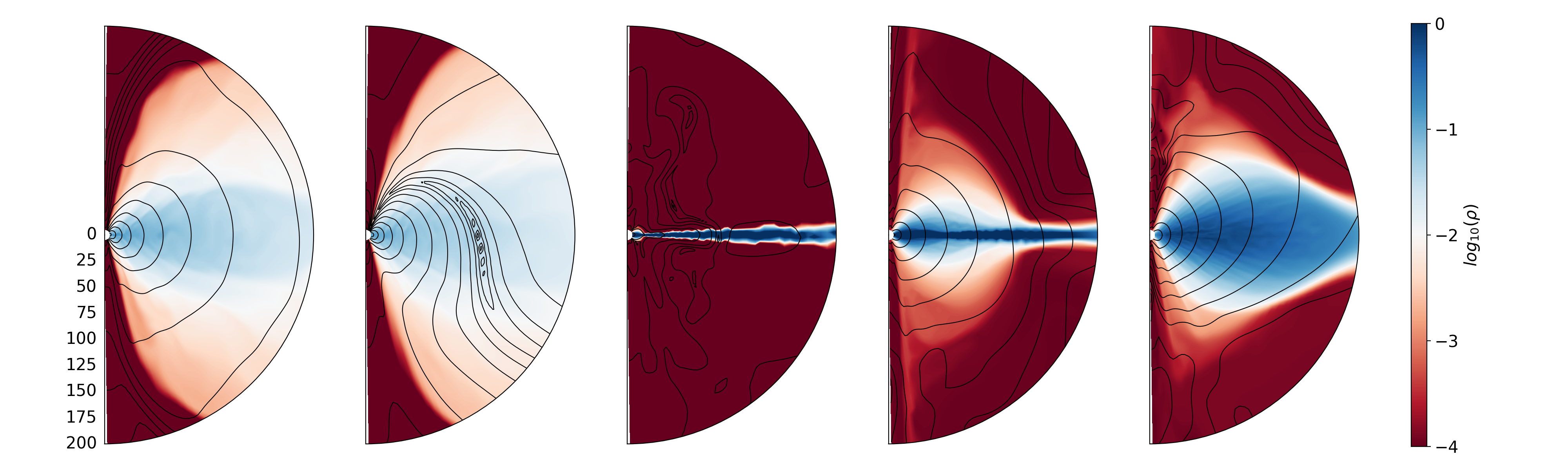}
    \caption{Density Contour plots of runs \texttt{AM.11.1}, \texttt{AM.10.1}, \texttt{AM.11.01}, \texttt{AM.10.01} and \texttt{AM.09.01} respectively with $\alpha_m=\mu=0.1$ after $t=100$ days. Black lines indicate magnetic contours. The first two runs have a magnetic Prandlt number ($\Pr_m$)$ = 0.667$ while $Pr_m=0.067$ in the next three runs. A comparatively thicker RIAF is observed for the first two plots, with the RIAF-ambient medium interface extending up to $200R_g$. In run \texttt{AM.10.1}, the magnetic skewed toroidal fields are not able to affect the accretion flow since the magnetic fields get diffused due to a higher value of resistivity.  The evidence of the corona is smaller in the run \texttt{AM.11.01}. Jet launching and magnetised outflows originating from the corona can be seen in \texttt{AM.10.01} and \texttt{AM.09.01} respectively due to well-ordered toroidal fields. The y range is in units of $R_g$.}
    \label{fig:alph_m-mag-contour}
\end{figure*}

\begin{figure*}
    \centering
    \includegraphics[width=1.0\linewidth]{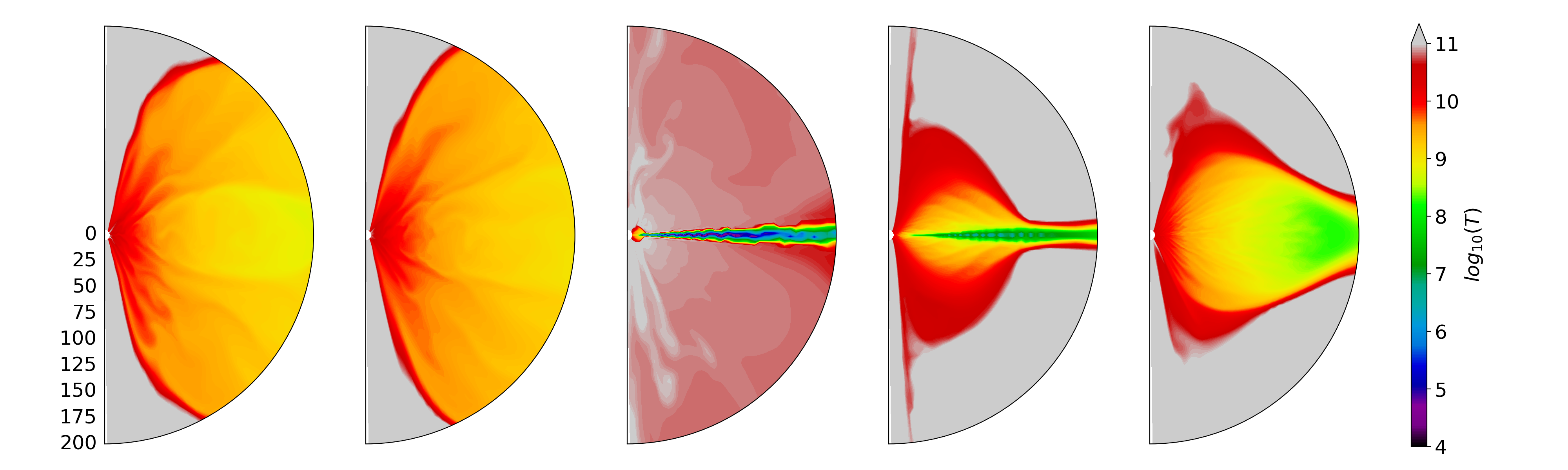}
    \caption{Temperature Contour plots of runs \texttt{AM.11.1}, \texttt{AM.10.1}, \texttt{AM.11.01}, \texttt{AM.10.01} and \texttt{AM.09.01} respectively with $\alpha_m=\mu=0.1$ after $t=100$ days. Again, the thin disk temperature decreases with $\rho_0$ and, consequently, the accretion rate. The global temperature of runs \texttt{AM.11.1} and \texttt{AM.10.1} is higher ($10^9-10^{10}K$) than the global temperature of \texttt{AM.09.01} ($10^8-10^9K$) due to a higher viscous heating rate, which is controlled by the prescription of $\alpha_v$. The y range is in units of $R_g$.}
    \label{fig:alph_m-temp}
\end{figure*}

\begin{figure*}
    \centering
    \includegraphics[width=1.0\linewidth]{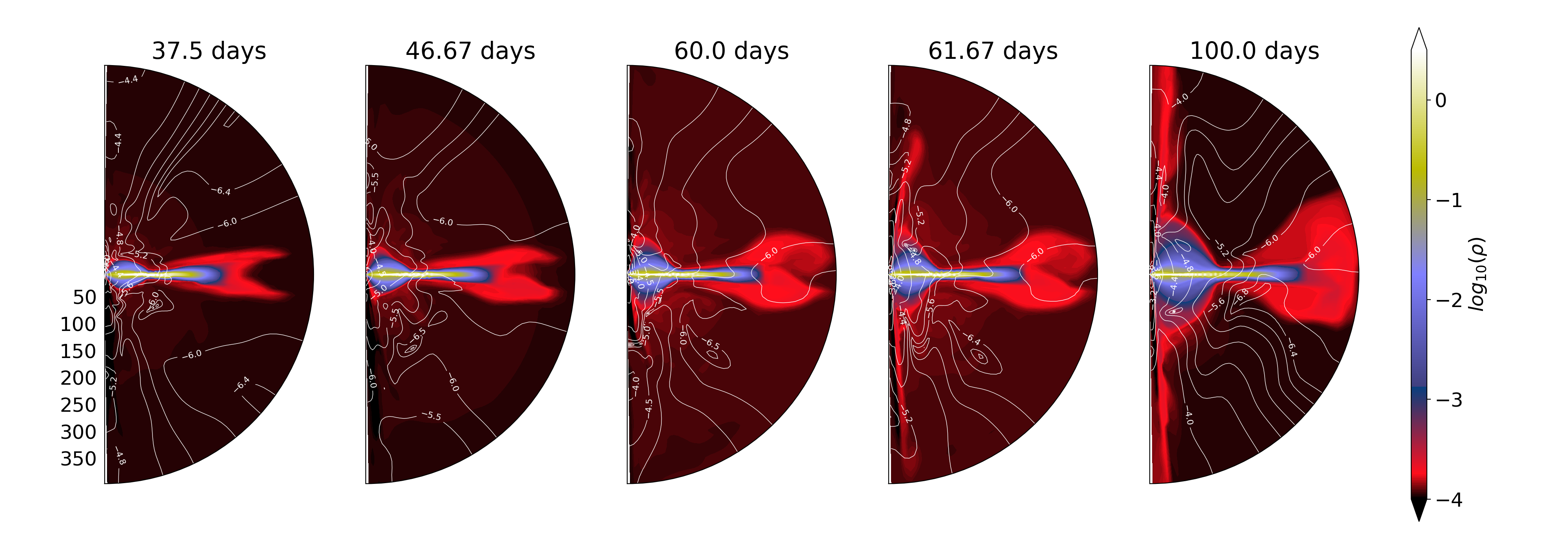}
    \caption{Persistent jet launching in run \texttt{AM.10.01}. Different contours and colour scaling have been used to visualise the jet launching properly, clearly showing the jet originating from the corona. White lines indicate magnetic contours in the $r-\theta$ plane, showing the presence of axial poloidal fields. The y range is in units of $R_g$.}
    \label{fig:alph_m-jet}
\end{figure*}

\begin{figure*}
    \centering
    \includegraphics[width=1.0\linewidth]{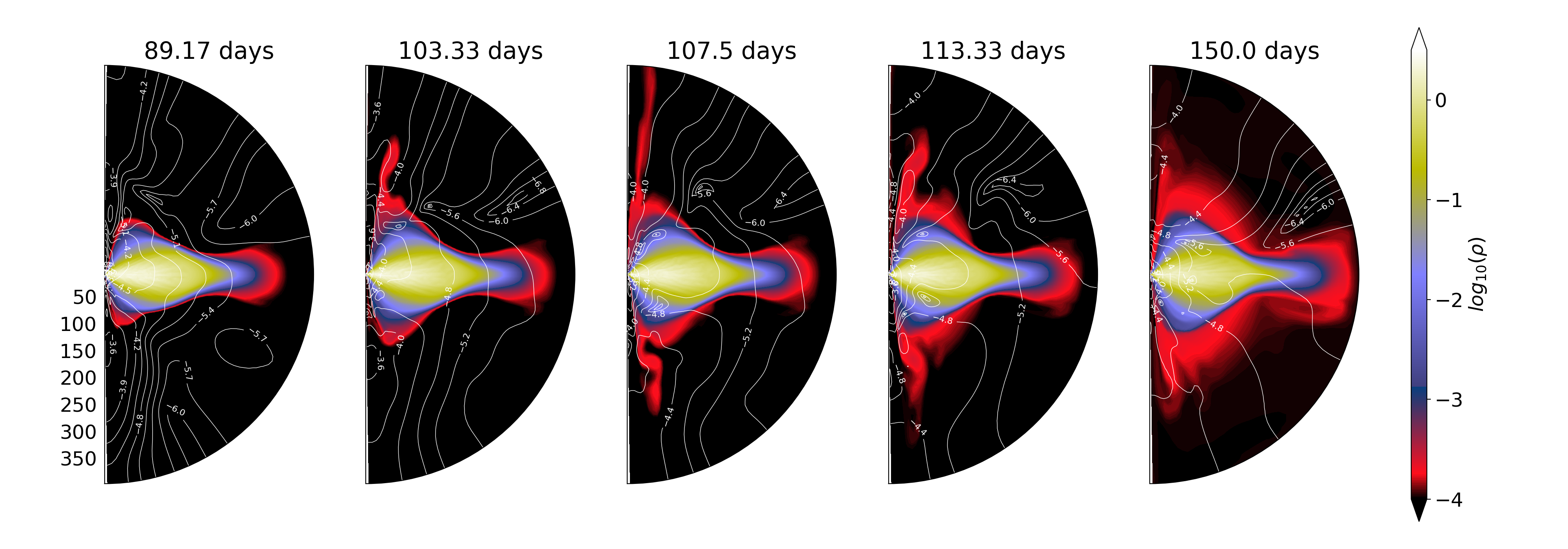}
    \caption{Magnetised outflow observed in run \texttt{AM.09.01}. White lines represent magnetic contours in the $r-\theta$ plane. The outflows originate from the hot accretion flow, which gets skewed due to the presence of toroidal fields. The y range is in units of $R_g$.}
    \label{fig:alph_m-outflow}
\end{figure*}

Next, we investigate whether the magnetic diffusivity parameter $\alpha_m$ has any effect on the transition for a constant value of the seed magnetic field. Fig. \ref{fig:alph_m-mag-contour} represents cases with $\mu=0.1$ and $\alpha_m=0.1$. For $\alpha_v=0.1$, no transition is observed, which is expected since a higher prescription of $\alpha_v$ implies a higher viscous heating rate. For $\alpha_v=0.01$, an initial density of $\rho_0=10^{11}\ m_p/cc$ produces similar results, but in this case, the RIAF extends vertically till $100R_g$, and the flow is comparatively denser. For $\rho_0=10^{10}\ m_p/cc$, a persistent jet is observed, which was not seen for $\alpha_m=0.01$ and is in accordance with \cite{kuwabara05}. In the case of the RIAF with a lower accretion rate ($\rho_0=10^{9}\ m_p/cc$), magnetized outflows are observed. Again, a higher accretion rate boasts a cooler thin disk (Fig. \ref{fig:alph_m-temp}). Run \texttt{AM.11.01} reports a thin disk with its midplane temperature raging from $10^{4}-10^{5}\ K$ while run \texttt{AM.10.01} reports a midplane temperature in the range $10^{5}-10^{6}\ K$. Thus, we notice that the temperature range is an order of magnitude less than their $\alpha_m=0.01$ counterparts and is close to the purely hydrodynamic runs. This confirms that resistive and reconnection heating increases the thin disk temperature. The aspect ratio profiles are similar to that observed in Fig \ref{fig:rho-accretion-rate}. However, run \texttt{AM.10.1} shows an accretion rate, which is an order of magnitude higher than runs with $\alpha_v=0.01$. Similar jumps have been observed in runs \texttt{AV.1.01} and \texttt{AV.1.1} (Sec. \ref{Res:alpha_v}). This is expected since an increase in $\alpha_v$ means a more efficient transfer of angular momentum outwards. This jump in accretion rate is not observed in run \texttt{AM.11.1} since an initial geometrically thin flow is observed, which gradually develops into a RIAF. Focusing on run \texttt{AM.10.01} (Fig \ref{fig:alph_m-jet}), we confirm that the jet is launched from the coronal material as the corona gets skewed toward the poles. The jets are launched at speeds of $2c/3$ and acquire a velocity close to  $c/2$ with time. Although this is not a relativistic simulation, the speeds will be very close to the speed of light, even for a proper relativistic simulation. Run \texttt{AM.09.01} (Fig. \ref{fig:alph_m-outflow}) does not show any persistent jet launching but rather occasional relativistic magnetised outflows with similar speeds. The jet launching is not observed for run \texttt{AM.11.01} for similar values of plasma beta and Magnetic Prandtl number ($Pr_m$) since the coronal part is much smaller (has a height of approximately $10\ R_g$), and we can conclude that jet is primarily launched from the hot corona and not the thin disk. Note: The jet launching has also not been observed for runs with the inner radial boundary limited to $15R_g$ for favourable conditions, indicating the requirement of stronger magnetic fields and higher angular velocities for jet launching, which are found very near the central object ($<10R_g$). Thus, we conclude that the proper relativistic jets can get launched under suitable conditions of magnetic field ($\beta<1000$) and $\alpha_m>$ ($Pr_m\leq0.05$) but we digress since this is beyond the scope of our present work and will be explored in much more details incorporating GRMHD simulations in future works.

\begin{figure*}
    \centering
    \includegraphics[width=0.49\linewidth]{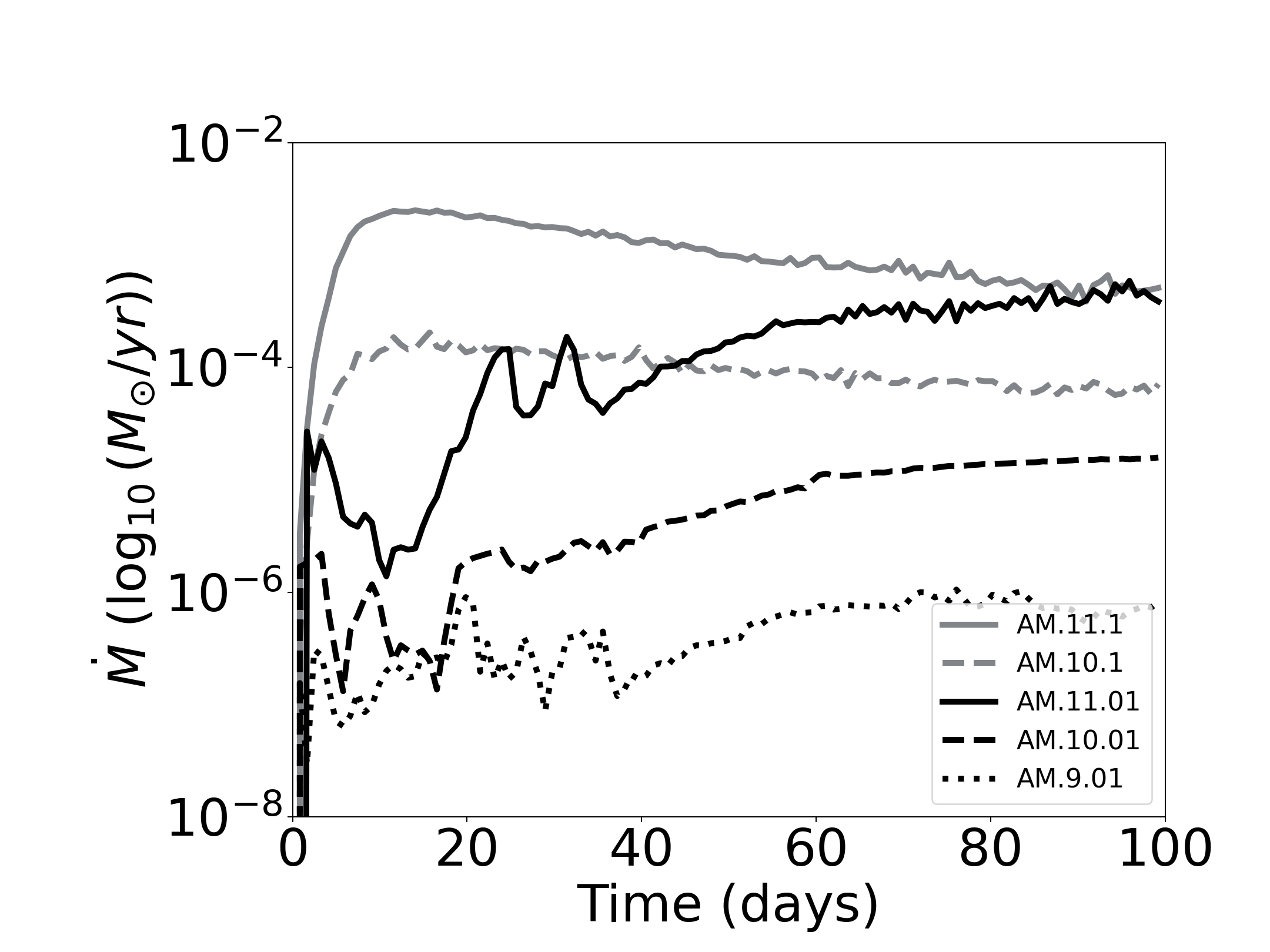}
    \includegraphics[width=0.49\linewidth]{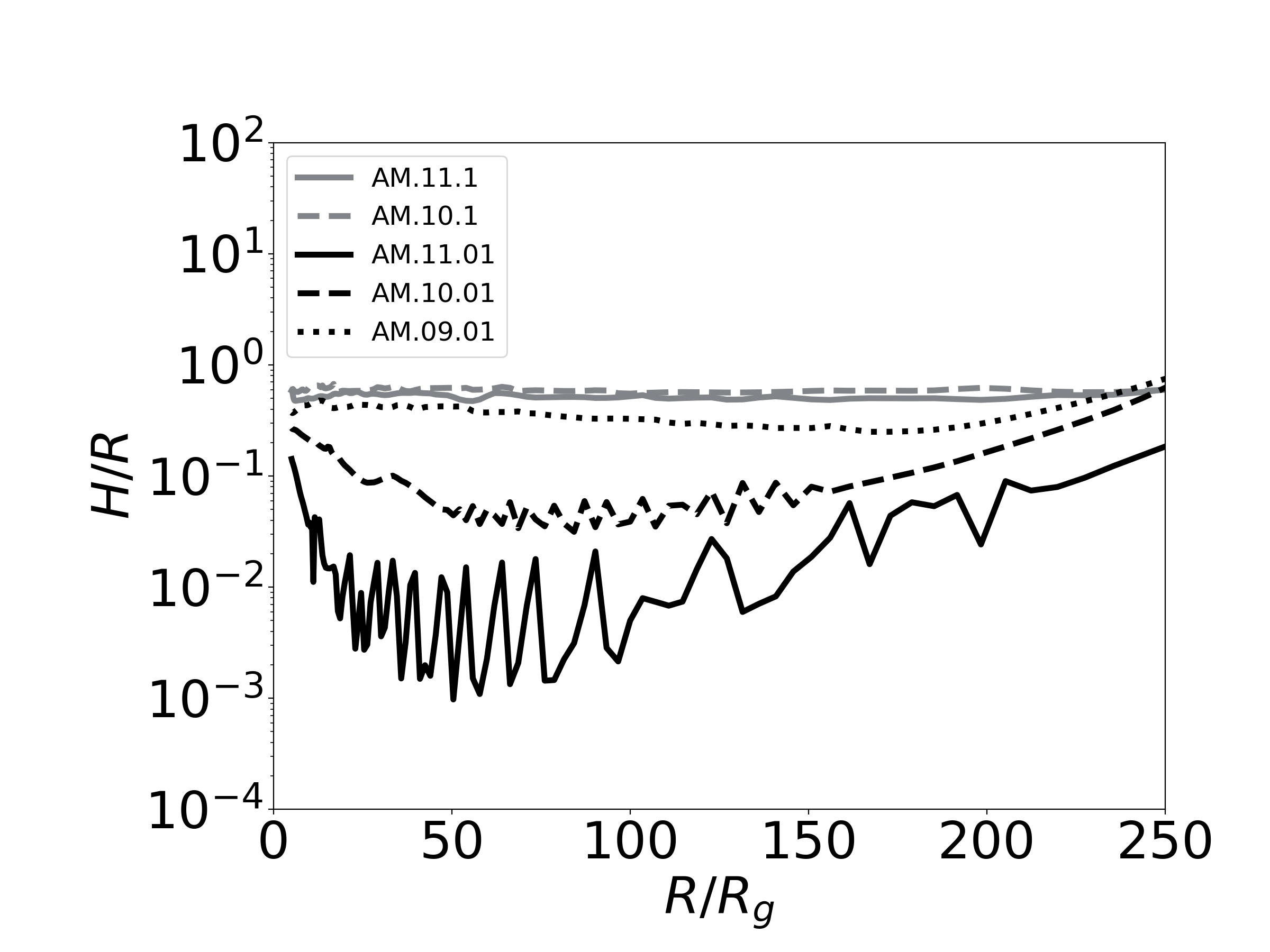}
    \caption{Left panel: Accretion rate vs time for runs with $\alpha_m=0.1$. \texttt{AM.10.1} has an accretion rate 1 order of magnitude higher than \texttt{AM.10.01} due to its higher value of $\alpha_v$ which controls the transport of angular momentum. Right panel: Disk midplane aspect ratios are similar to HD cases since the magnetic field is diffused due to an order of magnitude lower value of $Pr_m$ as opposed to runs with $alpha_m=0.01$.}
    \label{fig:alph_m-accretion-rate}
\end{figure*}

\subsection{\label{Res:alpha_v}Momentum diffusivity ($\alpha_v$)}

\begin{figure*}
    \centering
    \includegraphics[width=1.0\linewidth]{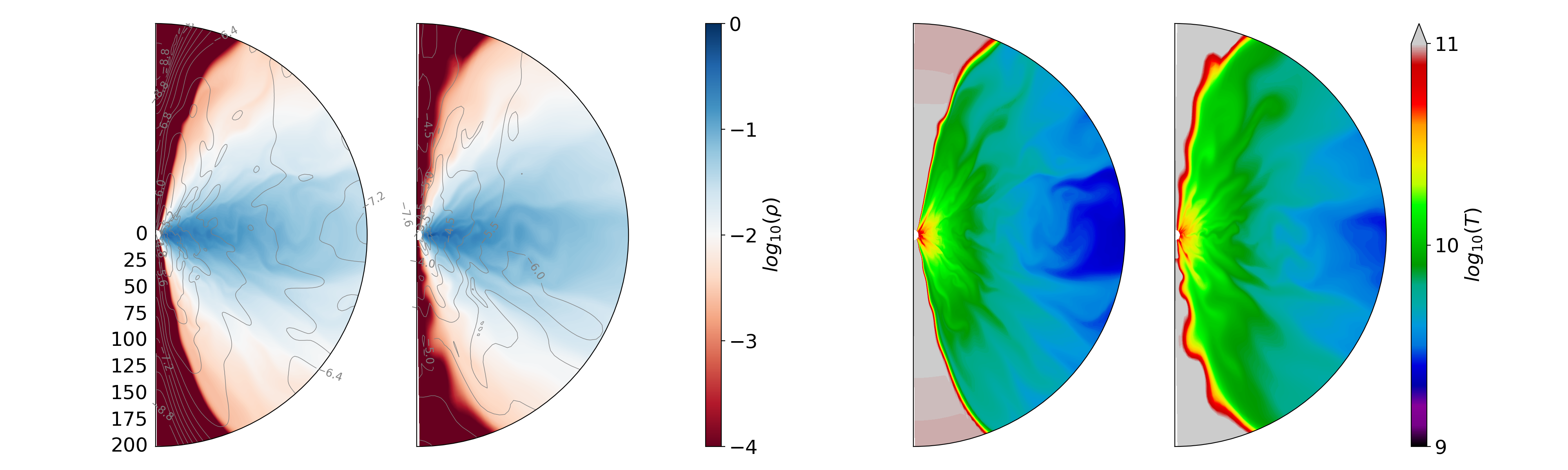}
    \caption{Left panel: Density contour plots of runs \texttt{AV.01} and \texttt{AV.1} respectively. Right panel: Density contour plots of runs \texttt{AV.01} and \texttt{AV.1} respectively. Both show similar accretion flow structures since it is primarily controlled by $\alpha_v$. Run \texttt{AV.1} is comparatively more turbulent since the magnetic field is an order of magnitude higher. The y range is in units of $R_g$.}
    \label{fig:alph_v-contours}
\end{figure*}

We have done runs with an $\alpha_v$ value of 0.1, which shows flaring (Fig. \ref{fig:alph_v-contours}). Run \texttt{AV.1} shows turbulence at the RIAF-ambient medium interface. The accretion rate and disk aspect ratio are similar in both runs, showing that these are independent of the specific seed magnetic field. The RIAF structure will change for lower values of plasma beta, but this is currently beyond the scope of this work. It is to be noted that the observed accretion rate at the quasi-steady state, in this case, is an order of magnitude higher than the $\alpha_v=0.01$ counterparts. Again, we highlight that this is primarily due to the more efficient transfer of angular momentum outwards.

\begin{figure*}
    \centering
    \includegraphics[width=0.49\linewidth]{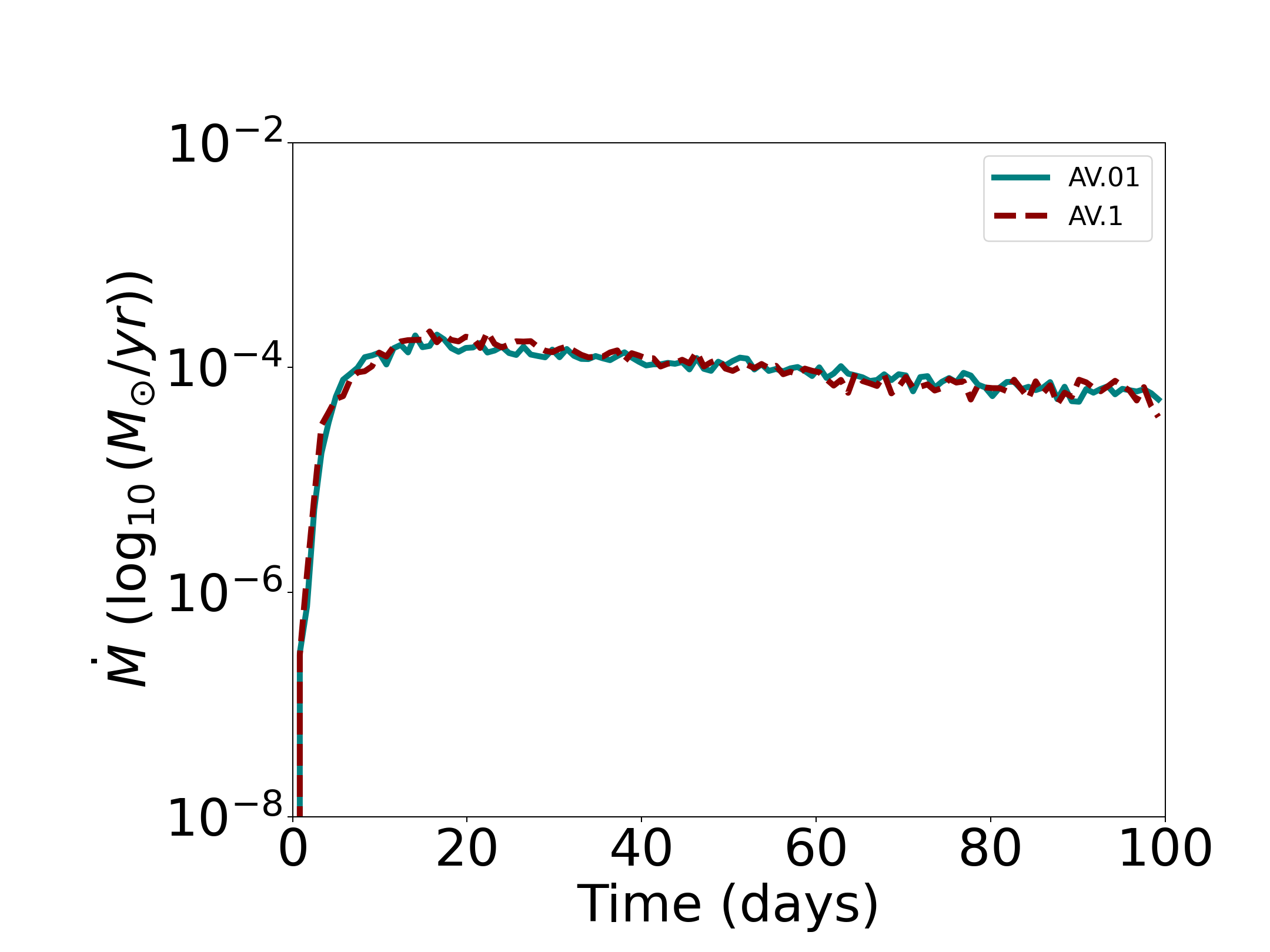}
    \includegraphics[width=0.49\linewidth]{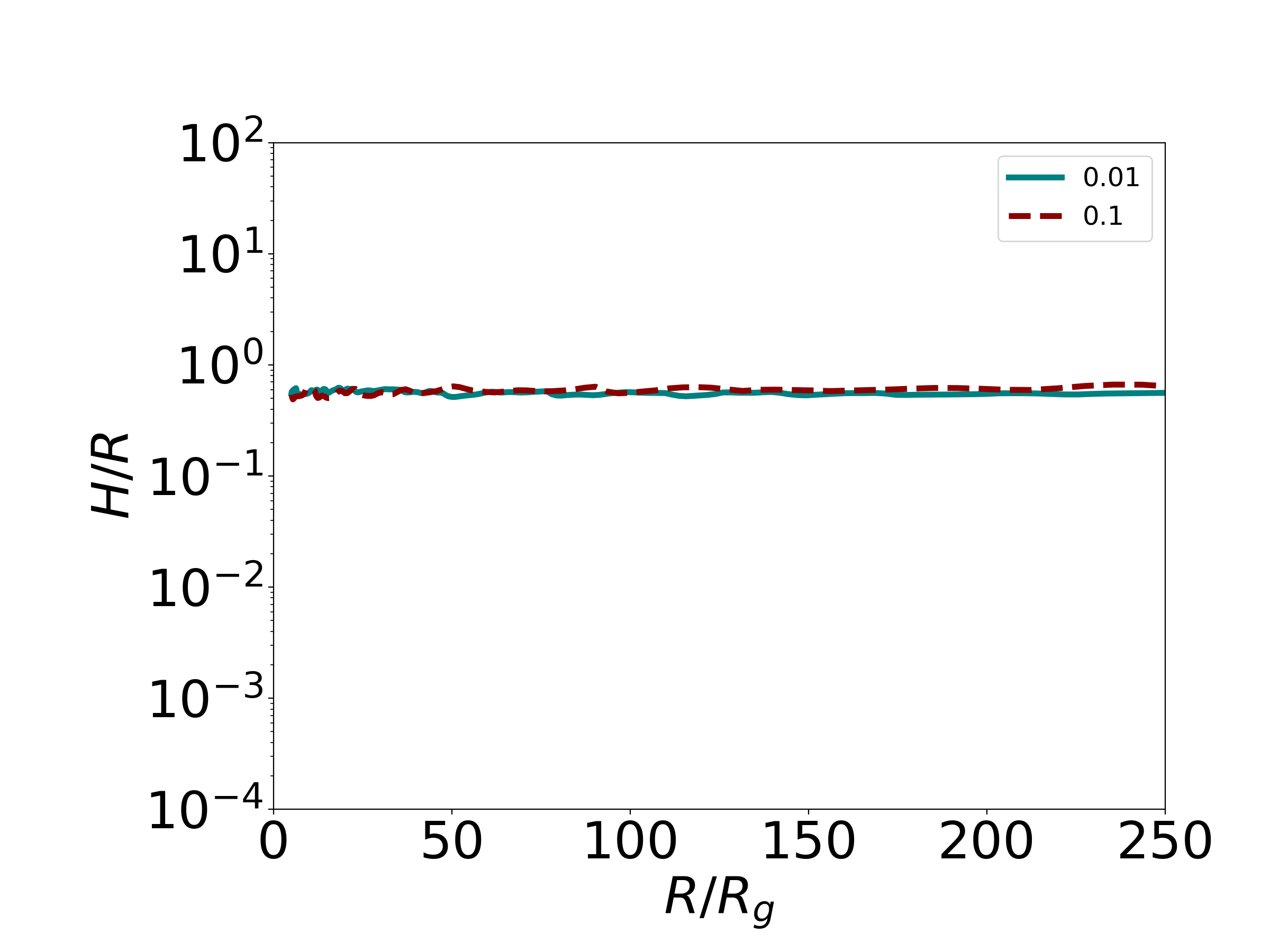}
    \caption{Left panel: Accretion rate for runs \texttt{AV.01} and \texttt{AV.1} respectively. The accretion rate in both cases is an order of magnitude higher than the $\alpha_v=0.01$ counterpart, indicating the influence of viscosity in momentum transport. Right panel: Aspect ratio profiles for runs \texttt{AV.01} and \texttt{AV.1} respectively shows that they are close to $1$. This is expected since no thin disk is formed.}
    \label{fig:alph_v-accertion-rate}
\end{figure*}

\section{\label{sec:Analysis}Analysis}

\begin{figure*}
    \centering
    \includegraphics[width=1.0\linewidth]{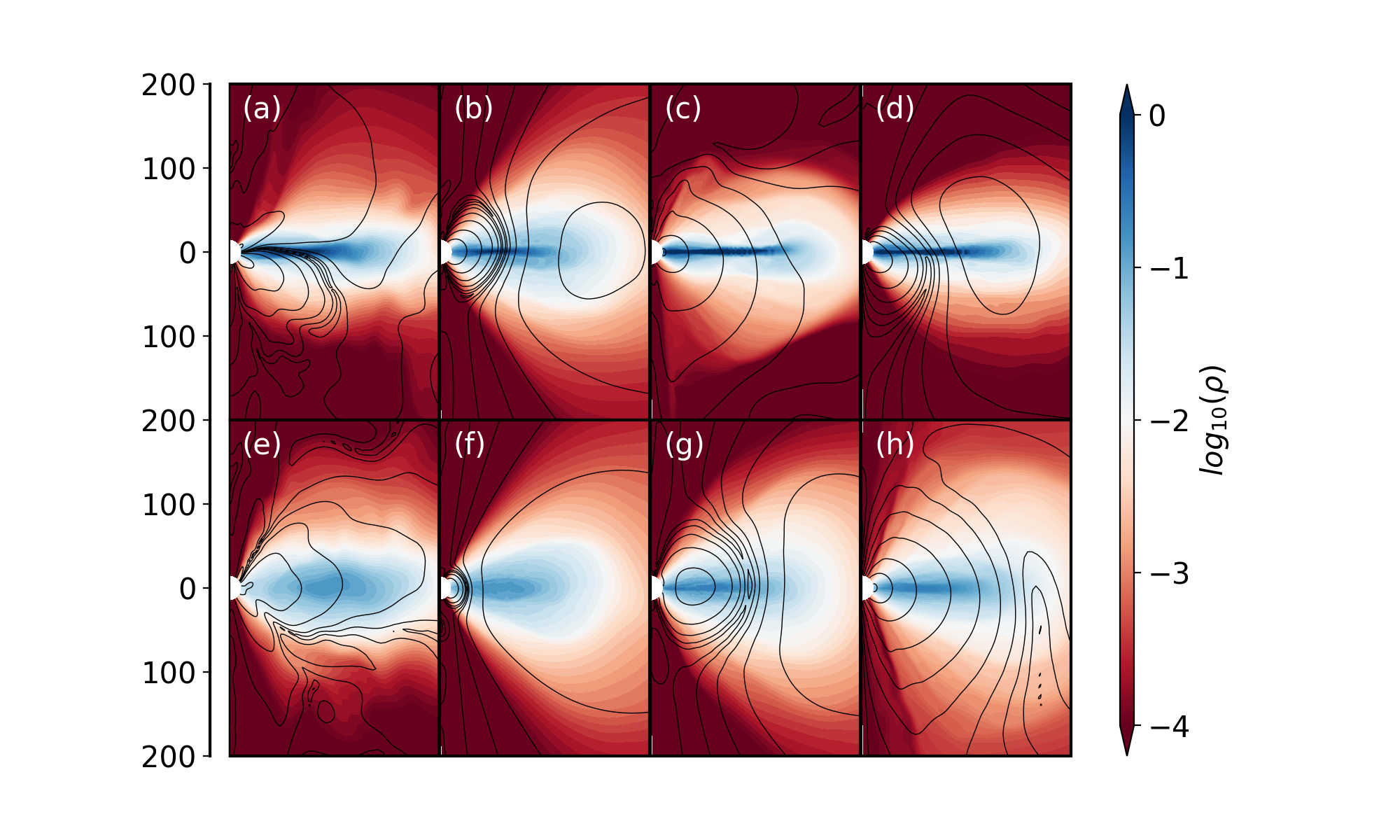}
    \caption{Density contours of runs \texttt{B.010.2}, \texttt{B.025.4}, \texttt{B.050.6}, \texttt{B.075.8} (Fig. (a), (b), (c) and (d) respectively) and  \texttt{B.010.3}, \texttt{B.025.5}, \texttt{B.050.7}, \texttt{B.075.9} (Fig. (e), (f), (g) and (h) respectively). The top panels show runs barely forming a thin disk. The bottom panels show an RIAF, which results in increasing $\mu$ by 0.1 in each case. The y range is in units of $R_g$.}
    \label{fig:tansition-density}
\end{figure*}

\begin{figure*}
    \centering
    \includegraphics[width=1.0\linewidth]{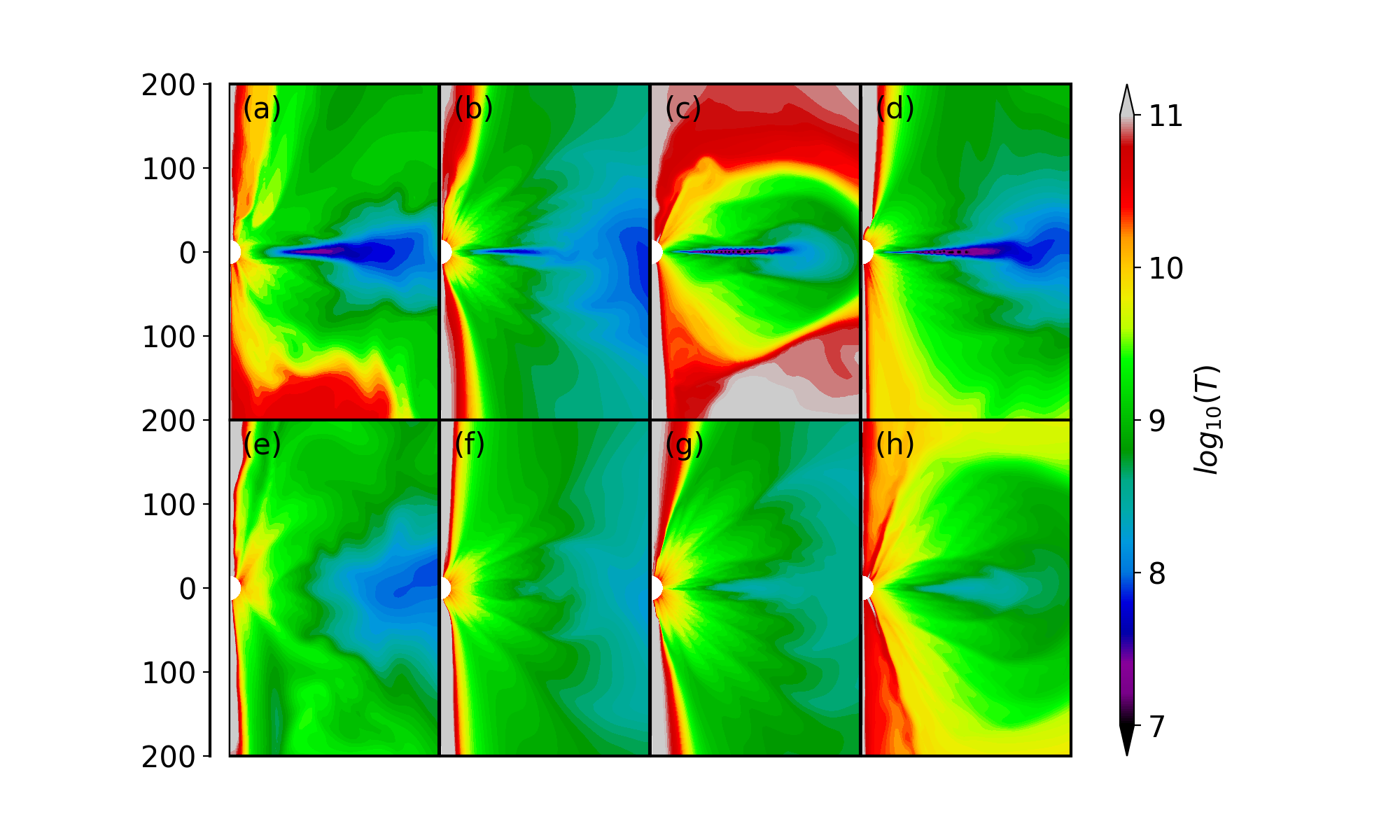}
    \caption{Temperature contours of runs \texttt{B.010.2}, \texttt{B.025.4}, \texttt{B.050.6}, \texttt{B.075.8} (Fig. (a), (b), (c) and (d) respectively) and  \texttt{B.010.3}, \texttt{B.025.5}, \texttt{B.050.7}, \texttt{B.075.9} (Fig. (e), (f), (g) and (h) respectively). The top panels show runs with a cool thin disk with a temperature in the range of $10^7-10^8K$. The bottom panel shows the RIAF with temperature $>10^8K$. The y range is in units of $R_g$.}
    \label{fig:tansition-temperature}
\end{figure*}

\begin{figure*}
    \centering
    \includegraphics[width=1.0\linewidth]{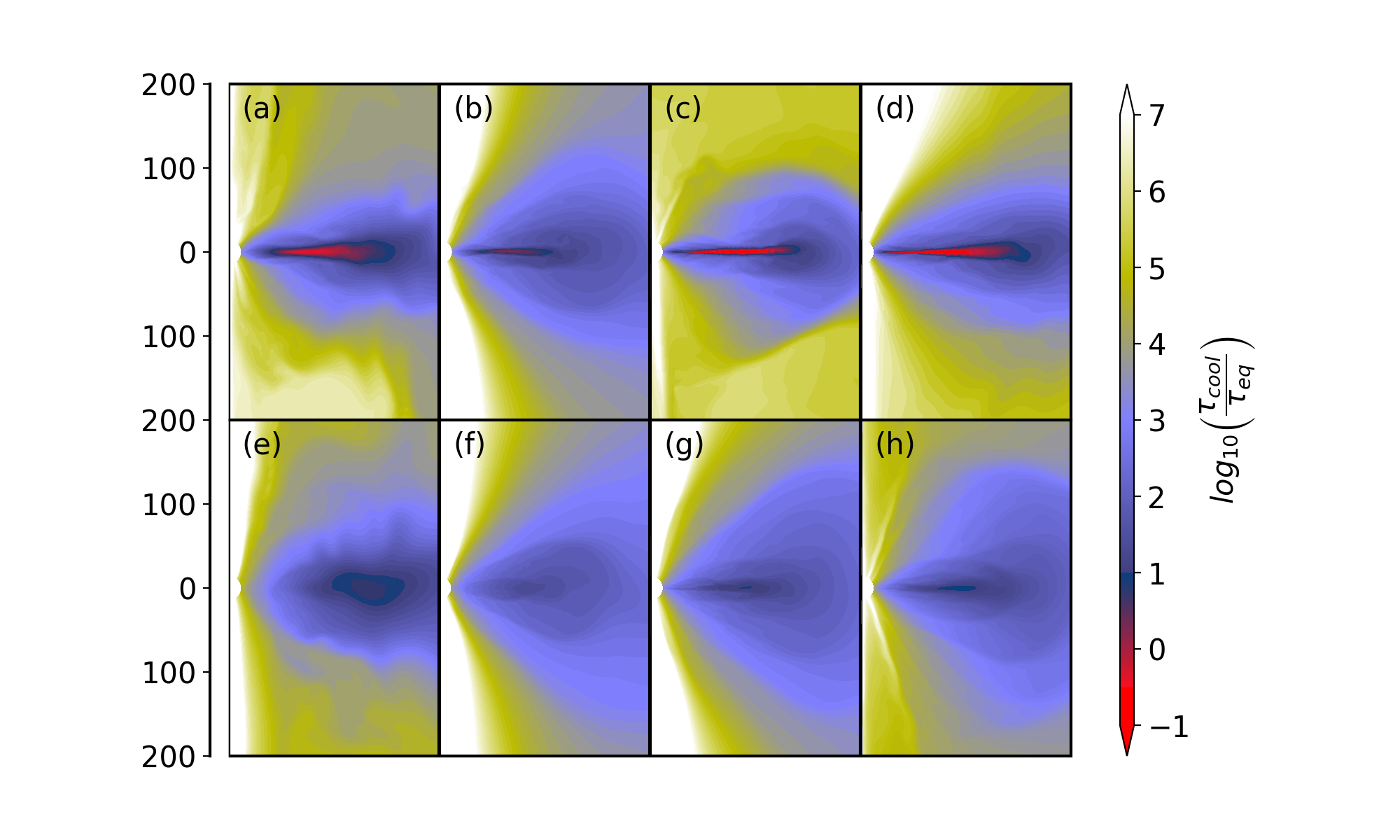}
    \caption{Contours of $\tau_{cool}/\tau_{eq}$ of runs \texttt{B.010.2}, \texttt{B.025.4}, \texttt{B.050.6}, \texttt{B.075.8} (Fig. (a), (b), (c) and (d) respectively) and  \texttt{B.010.3}, \texttt{B.025.5}, \texttt{B.050.7}, \texttt{B.075.9} (Fig. (e), (f), (g) and (h) respectively) where $\tau_{eq}$ is the harmonic mean of $\tau_{visc}$, $\tau_{res}$, $\tau_{recc}$ and $\tau_{MRIh}$. The cool thin disk only forms when $\tau_{cool}/\tau_{eq}<1$ (top panel: indicated by red contours) and agrees with Fig. \ref{fig:tansition-density} and Fig .\ref{fig:tansition-temperature}. The y range is in units of $R_g$.}
    \label{fig:tansition-time-scale}
\end{figure*}

\begin{figure*}
    \centering
    \includegraphics[width=0.49\linewidth]{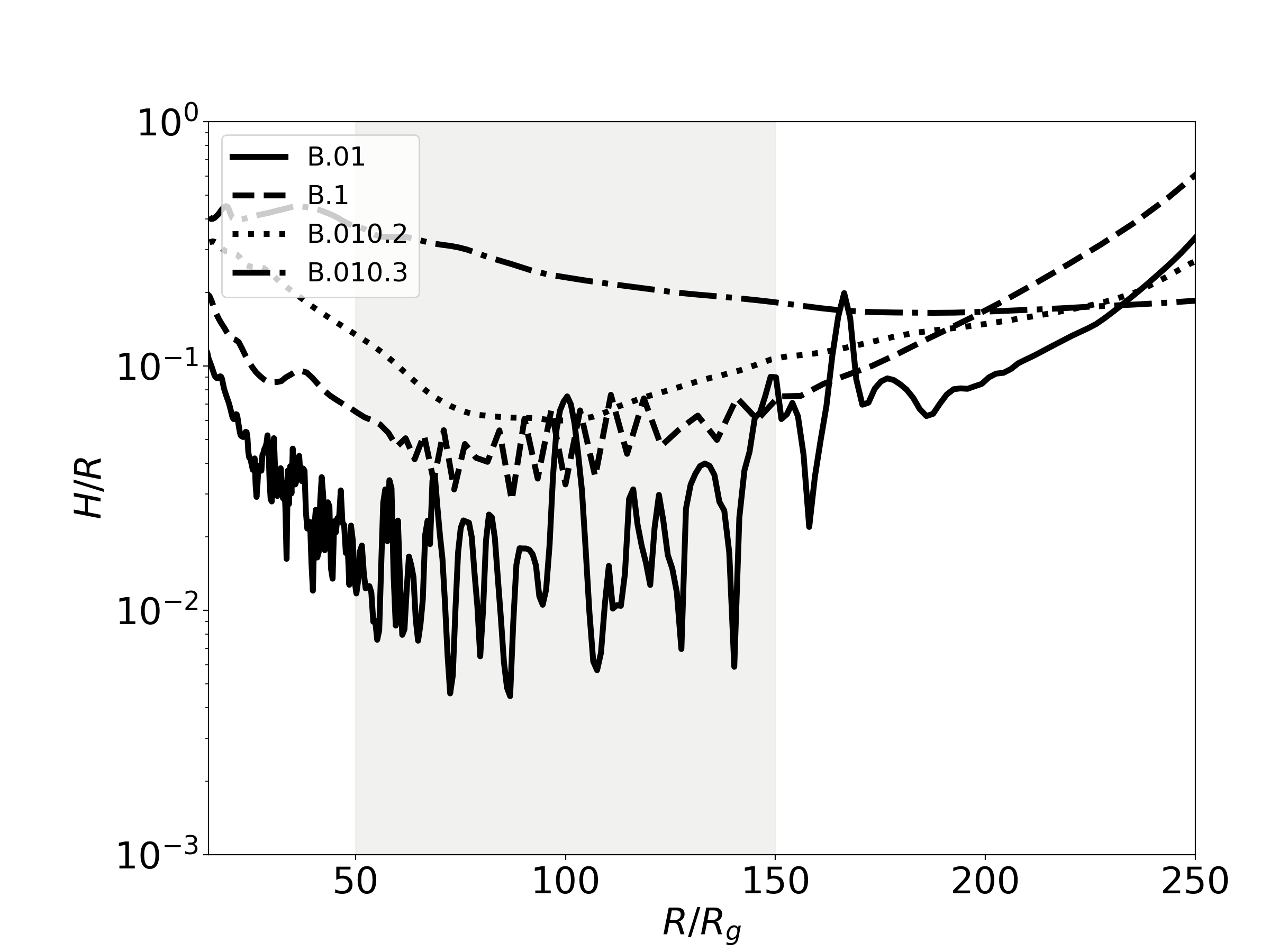}
    \includegraphics[width=0.49\linewidth]{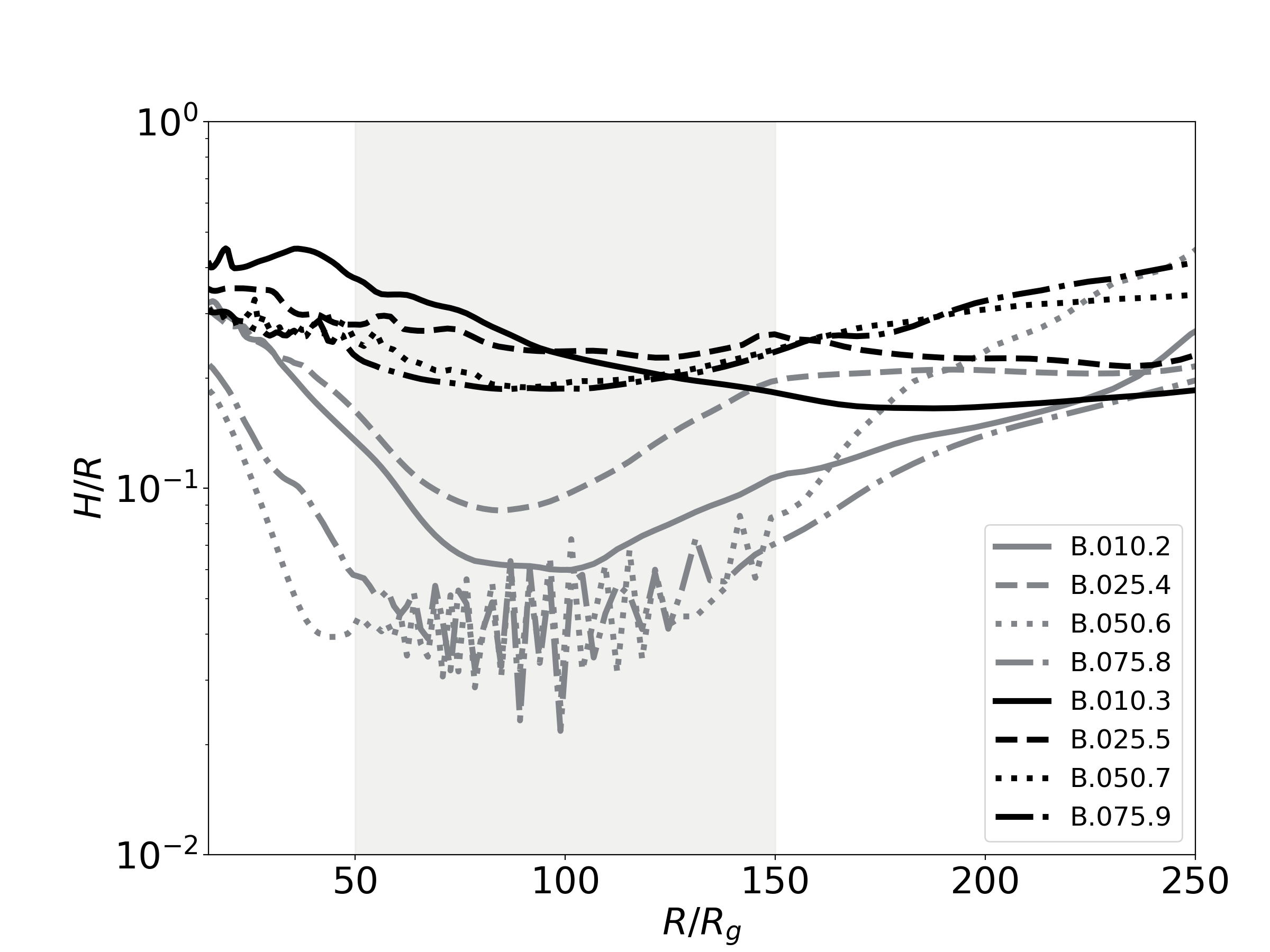}
    \caption{Left panel: Aspect ratio plots of runs \texttt{B.01}, \texttt{B.1}, \texttt{B.010.2} and \texttt{B.010.3}. The aspect ratio increases with an increase in the value of the seed magnetic field. Right panel: Aspect ratio of runs with $\mu$ difference of $0.1$ for equal $\alpha_m$ values. The runs with the lower value of $\mu$ have similar aspect ratios and are grouped together (dark grey lines), whereas runs where $\mu$ is greater than $0.1$ for the same value of $\alpha_m$ (black lines) are above $H/R=0.1$ and does not form a thin disk. The $50-100R_g$ region has been highlighted using light grey since the thin disk is present in this region in all runs. \textit{Note}: This region does not indicate the inner radius of formation of the thin disk or the outer radial extent of the thin disk.}
    \label{fig:transition-aspect-ratio}
\end{figure*}

\begin{figure*}
    \centering
    \includegraphics[width=0.49\linewidth]{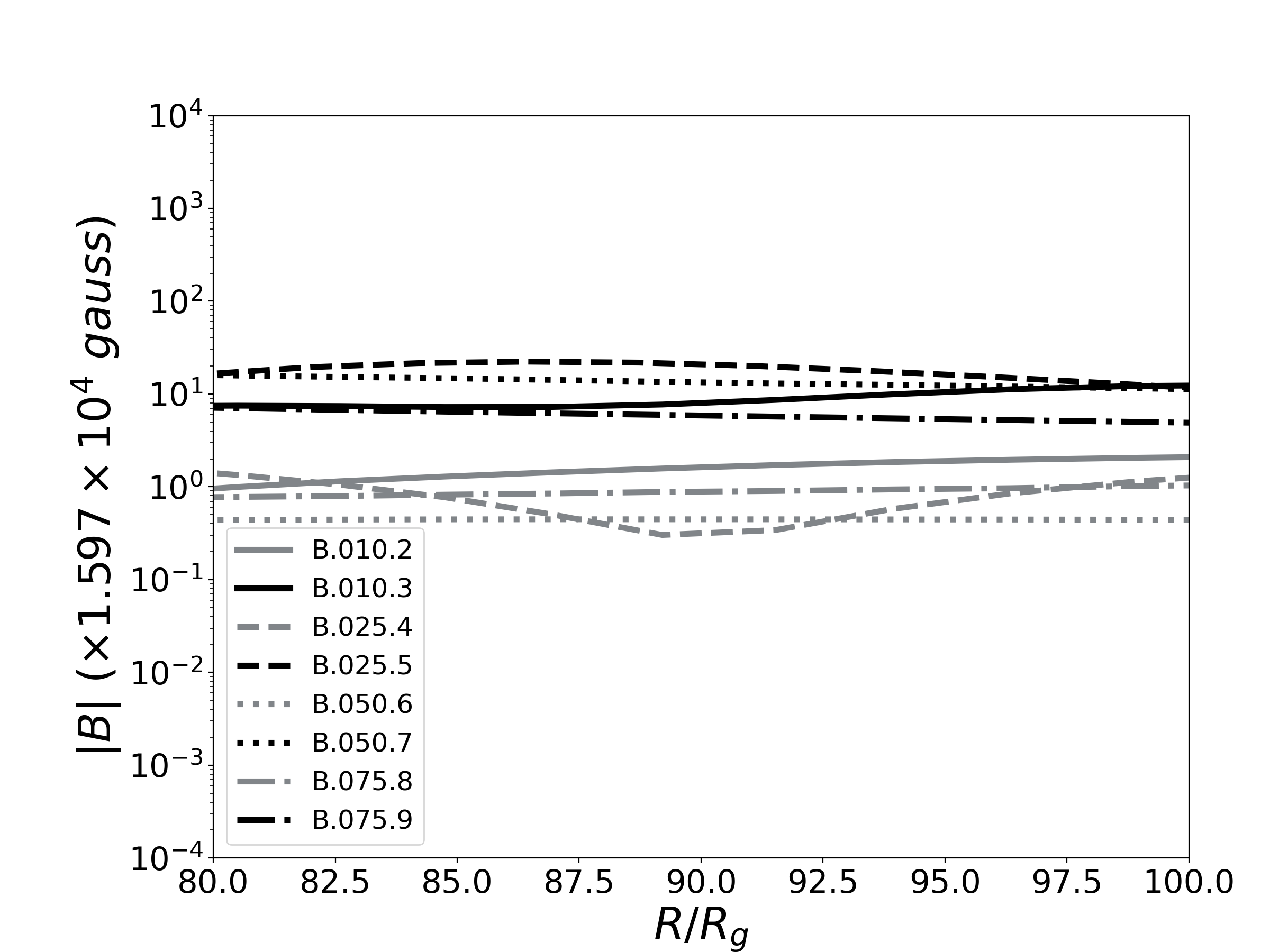}
    \includegraphics[width=0.49\linewidth]{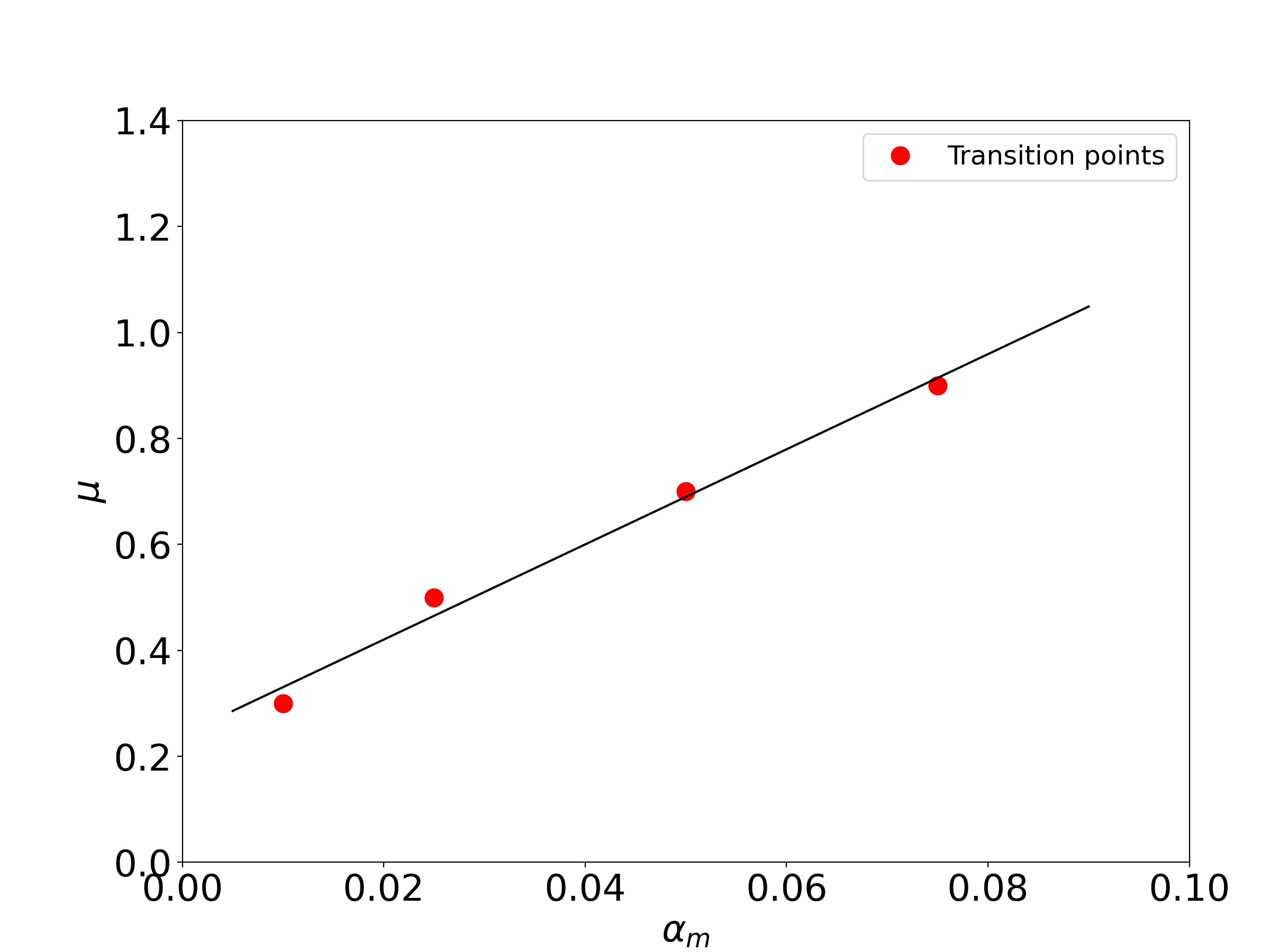}
    \caption{Left panel: Magnetic field strength (in code units) between $80-100R_g$ of runs where the thin disk exists (indicated in dark grey) as compared to runs where the thin disk does not exist for the same values of magnetic diffusivity (indicated in black). Despite the seed magnetic field being higher for higher $\alpha_m$ values, the RIAF is formed for the same order of magnetic field strength. The two cases are observed in the same order of magnitudes, separated by approximately an order of magnitude. The region of $80-100R_g$ has been selected since the magnetic field strength is comparatively uniform in this region. Right panel: A linear relationship exists between initial magnetic field strength controlled by $\mu$ and $\alpha_m$.}
    \label{fig:transition-magnetic-distribution}
\end{figure*}

To correctly model our results, we have to take into account the time scales of various sources of heating and cooling. The time scale of cooling has to be lower than an equivalent time scale of heating, which takes into account all the processes. We define the $\tau_{cool}$, $\tau_{visc}$, $\tau_{res}$, $\tau_{MRIh}$ and $\tau_{recc}$ as the typical timescales of cooling, viscosity, resistivity, MRI (Magneto-rotational Instability) heating and heating due to magnetic reconnection. For the purely hydrodynamic case, the transition occurs for the condition $\tau_{cool}/\tau_{visc}<1$ \cite{sharma2013}.

\begin{equation}
    \label{eqn:t-cool}
    \tau_{cool}=e/\dot{e}
\end{equation}

where $e$ is the internal energy density and $\dot{e}$ represents the cooling rate. The cooling rate is given as $\dot{e}=n_in_e\Lambda(T)\approx n^2\Lambda(T)$ and the internal energy density can be approximated as $e\approx 3nk_BT/2$ where $n$ represents the number density, $k_B$ is the Boltzmann constant and $T$ represents the local temperature. Thus, we have 

\begin{equation}
    \label{eqn:t-cool-mod}
    \tau_{cool}=\dfrac{3nk_BT}{2n^2\Lambda(T)}
\end{equation}

The viscous heating time scale is given as

\begin{equation}
    \label{eqn:t-visc}
    \tau_{visc}=\dfrac{r^2}{\nu_v}
\end{equation}

Similarly, the resistive heating timescale is defined as 

\begin{equation}
    \label{eqn:t-res}
    \tau_{visc}=\dfrac{r^2}{\nu_m}
\end{equation}

Next, we define the magnetic reconnection heating timescale. For our purpose, we consider the Sweet-Parker reconnection model \cite{parker63}. Although the Petschek reconnection model \cite{axford65} is capable of explaining more efficient energy loss, the diffusion length scales involved across shocks are much smaller than what can be resolved in our simulation except for very near the boundary \cite{kulsrud01}. Thus, for our analysis, we have only considered the Sweet-Parker reconnection model, which can occur due to toroidal current sheets driven by the toroidal magnetic fields, which are relatively stronger than the poloidal magnetic fields and also the reconnection of poloidal fields at the midplane. The typical timescale for the Sweet-Parker reconnection heating is a function of both the resistive timescale and the Alfven timescale and is given as 

\begin{equation}
    \label{eqn:t-recc}
    \tau_{recc}=\sqrt{\tau_{res}\tau_{alf}}
\end{equation}

where the Alfven timescale ($\tau_{alf}$) is defined by $r^2/v_A$, with $v_A=B/\sqrt{4\pi\rho}$ being the Alfven speed and $r$ being the typical length scale of such reconnection. Finally, we consider the MRI heating timescale. Since the angular velocity of the disk primarily controls the MRI growth rate, the MRI timescale can be approximated to be of the order of the orbital period. Although this is a crude approximation which does not take into account the strength of the magnetic field or the magnetic diffusivity coefficient, the order of the MRI heating timescale, which is primarily turbulence-driven, is typically higher than all the other timescales mentioned above since it is of the order of weeks and has been mentioned in our work purely for the sake of completeness. We approximate the MRI growth rate as $\gamma_{MRI}\approx\Omega$ and the MRI timescale as $\tau_{MRI}\approx1/\gamma_{MRI}\approx1/\Omega$. Using this, we can define the MRI heating timescale as 

\begin{equation}
    \label{eqn:t-MRI}
    \tau_{MRIh}=\dfrac{1}{\alpha_v\Omega}
\end{equation}

Having considered all the typical timescales, we need to define an equivalent timescale that is higher than the cooling time scale for the thin disk transition to be initiated. We note that the equivalent timescale $\tau_{eq}$ must be lower than the smallest heating time scale mentioned above since different sources of heating increase the heating rate. This motivated us to define the equivalent timescale as the harmonic mean of the fundamental heating timescales.

\begin{equation}
    \label{eqn:t-eq}
    \dfrac{1}{\tau_{eq}}=\dfrac{1}{\tau_{visc}} + \dfrac{1}{\tau_{res}} + \dfrac{1}{\tau_{recc}} + \dfrac{1}{\tau_{MRIh}}
\end{equation}

We assume that for the transition to occur from a low-hard state to a high-soft state, the following conditions must be satisfied:

\begin{equation}
    \label{eqn:transition-condition}
    \dfrac{\tau_{cool}}{\tau_{eq}}<1
\end{equation}

Eq. \ref{eqn:transition-condition} implies that The cooling rate must be higher than the net heating rate for such a transition to occur.\\
\\
Now we focus on our runs, where we observe that the RIAF is formed under 2 conditions:
\begin{enumerate}
    \item Increase in seed magnetic field strength for a constant value of magnetic diffusivity
    \item Further increase in seed magnetic field strength for increasing values of magnetic diffusivity
\end{enumerate}

Condition 1 indicates that magnetic reconnection plays a pivotal role in increasing the heating rate since a higher magnetic field would indicate higher Alfvenic speed and shorter Alfven time scales (Eq. \ref{eqn:t-recc}). Condition 2 indicates the effect of magnetic diffusivity or resistivity, which diffuses away the magnetic field. A higher seed magnetic field is required for magnetic pressure to evolve to the same order as seen in low diffusivity runs. However, we point out that the value of $\alpha_m$ depends on the chemical composition. From all these considerations, we understand that for lower magnetic field strengths ($\beta>22000$ for $\dot{M}\approx10^{-5}M_{\odot}/year$) being advected from the binary companion, the condition for the formation of the thin disc is similar to that of the hydrodynamic case and is correctly described as $\tau_{cool}/\tau_{visc}<1$. However, for advected fields with higher strengths, our model (Eq. \ref{eqn:transition-condition}) describes the flow correctly, and a thin disk may not be observed at all if the advected magnetic field strength is high enough. The left panel in Fig. \ref{fig:transition-aspect-ratio} shows a consistent increase in the aspect ratio with an increase in $\mu$ in the region of $50-100R_g$ where the part of the thin disk is present in all runs. The same region has been highlighted in the right panel, where we see that aspect ratios of runs with a lower value of $\mu$ (indicated in green) for corresponding values of $\alpha_m$ are separated by approximately an order of magnitude for the higher value of $\mu$ (indicated in black) where the thin disk is no longer present. The final magnetic field strength after 100 days (Fig. \ref{fig:transition-magnetic-distribution}) in both cases is again separated by an order of magnitude. This tells us that the transition occurs at a particular value of plasma beta for a specified accretion rate, and the increase in seed magnetic field strength counteracts the increase in magnetic diffusion. The seed magnetic field strength increases linearly with $\alpha_m$ (Fig. \ref{fig:transition-magnetic-distribution}). The thin disk lies in the temperature range of $10^{7}-10^{8}K$ (Fig. \ref{fig:tansition-temperature}). The expected thin disk, according to our model, is correctly able to explain the simulation results (Fig. \ref{fig:tansition-time-scale}).

\section{\label{sec:Conclusion}Discussions \& Conclusion}

In this study, we investigated the numerical evolution of both magnetized and non-magnetized, radiatively inefficient accretion flows within a parametric framework to examine how magnetic fields influence the formation of thin disks. To explore this, we varied the strength of the seed magnetic field and the magnetic diffusivity and analyzed their impact on the flow characteristics. We see an increase in thin disk temperature with an increase in initial seed magnetic field strength. This increase in temperature with an increase in seed magnetic field strength can be attributed to a higher reconnection rate as a result of higher Alfven speeds. Our findings reveal a critical seed magnetic field value of $\sim$52 gauss at an accretion rate of \(10^{-5}\ M_{\odot}/\text{year}\) and $\alpha$ magnetic diffusivity parameter ($\alpha_m$) of 0.03. Beyond this limit, the formation of a thin disk is not favourable, as indicated by our simulations and shown below. The above result can be attributed to the increased heating rate caused by the magnetic fields. To address the excess heating rate, we have introduced a new time scale, \(\tau_{eq}\), which incorporates the time scales of heat sources involving viscosity, resistivity, magnetic reconnection and MRI (Eq. \ref{eqn:t-eq}). We propose that for a thin disk to exist, this time scale must exceed the cooling time scale. By comparing the threshold values from our model with simulation results, we validate that our model accurately describes the behaviour of such accretion flows.

\subsection{Estimation of the threshold value of plasma beta}
We have estimated a threshold value for plasma beta, below which a geometrically thin accretion flow cannot be sustained. Since we have considered $\alpha$ parametrization for both resistivity and viscosity, both the timescales will be similar to each other (see Eq. \ref{eqn:nu} and Eq. \ref{eqn:nu_m}). We propose that the timescale of reconnection must be equal to or lower than that of the resistive timescale to affect the nature of the flow. The cooling time scale would counter a time scale of higher order. Thus using Eq. \ref{eqn:t-res}, and Eq. \ref{eqn:t-recc}, we propose:

\begin{equation}
\begin{split}
    \label{eqn:timescale-condition}
    \tau_{recc}&\approx\tau_{res}\\
    B&\approx4\pi\eta\sqrt{\pi\rho}
\end{split}
\end{equation}

We see that the threshold magnetic field strength $B$ is proportional to $\eta$, which is what is depicted in Fig. \ref{fig:transition-magnetic-distribution}. Substituting the scaled value of $\eta=1.5\times10^{-4}$ and $\rho=\rho_0=1$ at the distance of $100R_g$ for the initial condition yields a value of $0.00333$ which is approximately the value of initialization magnetic field strength which we have used in our simulation $\mu_{thresh}/r=0.003$ at a distance of $100R_g$. This leads us to define a threshold value of plasma beta as a function of density and magnetic diffusivity. Using Eq. \ref{eqn:timescale-condition}, plasma beta: defined as the ratio of thermal pressure to the magnetic pressure ($8\pi P/B^2$) for a polytropic equation of state becomes:

\begin{equation}
    \label{eqn:plasma-beta-condtn}
    \beta_{th}=\dfrac{K\rho^{\gamma-1}}{2\pi^2\eta^2}
\end{equation}
\\
Equation \ref{eqn:plasma-beta-condtn} provides a means to estimate the plasma beta beyond which a thin disk will be observed. If the material being transferred from a companion star onto a central object obeys a polytropic equation of state has a value of $\beta$ equal to or lower than the threshold provided by Eq. \ref{eqn:plasma-beta-condtn}, it will only form a RIAF and the geometrically thin, optically thick disk will not be observed. This $\beta$ is a function of position and a local value of $\beta$ greater than this threshold can give rise to local collapses. However, this interpretation may be misleading. The correct way to view this is by considering the value of $\beta$ at a position far away from the compact object ($100R_g$ in our case). Since both resistivity, viscosity and pressure are varying (as a function of $r$ in our case), the values at that particular distance is to be considered. If the $\beta$ value is less than the threshold at this position, the evolution of the accretion flow will end up in a RIAF rather than forming a thin disk. In our case, $\rho\approxprop (1/r)^{1/\gamma-1}$ from Eq. \ref{eqn:rho} and $\eta\propto \rho r^{1/2}$. Using these conditions and Eq. \ref{eqn:plasma-beta-condtn}, the $r$ dependence of $\beta$ in our case is given by

\begin{equation}
    \label{eqn:plasma-beta-condtn-particular}
    \beta_{th}(r)\propto r^{\left[\dfrac{\gamma+1}{2(\gamma-3)}\right]}
\end{equation}

Eq. \ref{eqn:plasma-beta-condtn-particular} shows the non-linear dependence of $\beta$ on the coordinate $r$. Thus the threshold is not a constant one but varies according to the viscosity coefficient and density which depends on $r$ in our case. From our simulations, the value of $\beta$ is $622$, while our model predicts a threshold beta value of $757$, which is of a similar order of magnitude. The slight difference in the value of $\beta$ can be attributed to numerical artifacts such as numerical diffusion. This is in accordance with observation where the source spends most of its time in the low-hard state rather than the high-soft state \cite{Homan04}. Alternatively, if the magnetic field gets amplified by the dynamo process due to an MRI, the excess heat generated in the inner region of the disk due to resistivity and reconnection at the midplane can cause a transition from a thin to thick accretion flow.

Due to our current limitation in computational resources, we are not able to carry out our simulations in higher resolution grids, extend our computational domain beyond $5R_g$ for high $\beta$ values ($>1000$) and $15R_g$ for comparatively lower $\beta$ values or take into account addition physical effects such as general relativity and radiation. These concerns will be addressed in future work.

\begin{acknowledgments}
AS acknowledges the usage of a local workstation server and PARAM Seva supercomputing facility hosted by the Indian Institute of Technology, Hyderabad. We thank Dr. Kirit Makwana for his suggestions and Dr. Dipanjan Mukherjee for helpful discussions and for hosting AS at IUCAA, Pune, for research discussions.
\end{acknowledgments}


\providecommand{\noopsort}[1]{}\providecommand{\singleletter}[1]{#1}%

\end{document}